\documentclass[aps,pre,groupedaddress,showpacs]{revtex4-1} 

\usepackage{graphicx}
\usepackage{subfigure}
\usepackage{amsmath}
\usepackage{amsfonts}
\usepackage{amssymb}
\usepackage{color}
\begin{document}

\title{Surface waves in granular phononic crystals} 
\author{H. Pichard}
\email[helene.pichard@univ-lemans.fr]{}

\author{A. Duclos}
\author{J-P. Groby}
\author{V. Tournat}
\author{L. Zheng}
\author{V.E. Gusev} 
\affiliation{LAUM, UMR-CNRS 6613, Universit\'e du Maine, Av. O. Messiaen, 72085 Le Mans, France}
\email[vitali.goussev@univ-lemans.fr]{}

\raggedbottom
\date{\today}

\begin{abstract}
The existence of surface elastic waves at a mechanically free surface of granular phononic crystals is studied. The granular phononic crystals are made of spherical particles distributed periodically on a simple cubic lattice. It is assumed that the particles are interacting by means of normal, shear and bending contact rigidities. First, Rayleigh-type surface acoustic waves, where the displacement of the particles takes place in the sagittal plane while the particles possess one rotational and two translational degrees of freedom, are analyzed. Second, shear-horizontal-type waves, where the displacement of the particles is normal to the sagittal plane while the particles possess one translational and two rotational degrees of freedom are studied. The existence of zero-group velocity surface acoustic waves of Rayleigh-type is theoretically predicted and interpreted. A comparison with surface waves predicted by the Cosserat theory is performed, and its limitations are established.
\end{abstract}

\pacs{68.35-iv,45.70-n, 63.20-e}

\maketitle

\section{Introduction}
The study of surface elastic/acoustic waves (SAWs) associated with the surface of a semi-infinite phononic crystal has attracted a lot of attention in recent years. The control of evanescent waves in periodic composites, both in photonic and phononic crystals, is promising for the design of new electromagnetic and acoustic materials for various applications~\cite{Tanaka98}. 
 An elementary volume of the medium possesses two translational degrees of freedom with the mechanical displacement vector polarized in the sagittal plane. To understand the effect of a free surface on the normal vibration modes of a crystal, continuous and discrete models have been applied to structures in one (chain), two (membrane) and three (half-space) dimensions, possessing different types of inter-atomic interactions. Investigations of surface modes of vibration using the continuum point of view have been reported for cubic crystals by Stoneley~\cite{Stoneley55} and by Gazis, Herman, and Wallis~\cite{Gazis60}. A description of surface waves for discrete lattices has been given by Lifshitz and Pekar~\cite{Lifshitz55}. Calculations based on specific lattice models have been given by Gazis and al.~\cite{Gazis60} for diatomic one-, two-, and three-dimensional NaCI-type lattices with nearest neighbor interactions only, and by Kaplan~\cite{Kaplan62} for the monatomic one-dimensional lattice with nearest and next-nearest neighbor interactions. Gazis, Herman, and Wallis treated the semi-infinite three-dimensional monatomic cubic lattice with nearest and next-nearest neighbor central forces and with angular stiffness forces. For long-wavelengths compared to the inter-atomic distance, the discrete particle theory, as to be expected, yields identical results to those of the continuum theory. When the wavelength becomes comparable to the interatomic distance the particle theory leads to dispersion, while the continuum results remain nondispersive for all wavelengths. \\

There are situations when the medium behavior is still elastic but the wave propagation cannot be described by the classical continuum elasticity theory because this theory does not properly describe the dispersion of the propagating long acoustic waves. To address this problem, polar (couple/asymmetric stress) elastic theories introduce supplementary and independent rotational degrees of freedom (\textit{dofs}) of material particles, which are additional to translational \textit{dofs} in classical continuum elasticity. Various models of this kind are widely used in continuum mechanics: Cosserat theory~\cite{Nowacki86}, micropolar model of Eringen~\cite{Eringen99}, reduced Cosserat continuum model~\cite{Grekova09}, etc. However, the lack of information on the additionally introduced physical parameters values for real materials hinders the development of these theories and their practical applications. In the Cosserat theory, each material element possesses six \textit{dofs}: three \textit{dofs} for the translation and three \textit{dofs} for the rotation. The Cosserat continuum elasticity theory predicts strong modification of the shear waves dispersion by the rotational \textit{dofs}. One of such effects is the dispersion of the Rayleigh surface elastic wave at flat interface for long-wavelengths, while the classical theory fails to explain it~\cite{Eringen99,Grekova09,Kulesh06,Kulesh06bis}. Furthermore, the Cosserat model predicts the propagation of horizontally polarized transversal surface waves~\cite{Kulesh07_b}, which are forbidden at the surface of the homogeneous classical elastic continuum.\\

Recently, discrete lattice models have been developed to describe the phononic band structure of granular crystals possessing rotational \textit{dofs}~\cite{Merkel09,Merkel10,Tournat11,Merkel11,Pichard12,Pichard14}. These discrete models provide elastic eigenmodes for all possible wavelengths. A two-dimensional (2D) discrete lattice model with particles possessing one translational and two rotational \textit{dofs} has been applied to the analysis of a monolayer granular phononic membrane~\cite{Tournat11}. It was demonstrated theoretically that the interaction between translational and rotational motions could lead to the opening of the gaps forbidden for wave propagation, to the creation of Dirac cone, to the existence of zero-energy soft modes and zero-group-velocity bulk modes~\cite{Pichard12}. Existence of localized modes have been demonstrated theoretically in a one-dimensional (1D) monatomic granular phononic crystal composed of infinitely long cylinders with equal masses and possessing one translational and one rotational \textit{dofs}~\cite{Pichard14}. Each of the localized coupled transversal and rotational mode existing in this studied chain is composed of two evanescent modes and is analyzed for different boundary conditions applied at the boundary of the semi-infinite chain. The experimental observation of the coupled rotational-translational bulk modes in a noncohesive granular phononic crystal was reported in~\cite{Merkel11}. It was demonstrated that the Cosserat theory generally fails to correctly predict the dispersion relations of the bulk elastic modes in granular crystals even in the long-wavelength limit because it does not account for all the effects of the material inhomogeneity. It should be combined with higher gradient elasticity theories~\cite{Eringen99,Kunin83,Muhlhaus95}.\\ 

This work focuses on the existence of SAWs at mechanically free surface of granular phononic crystals made of spherical particles. Rayleigh-type SAWs, where the particles possess one rotational and two translational \textit{dofs}, and shear-horizontal-type SAWs, where particles possess one translational and two rotational \textit{dofs}, are studied. Our analysis shows that the existence of SAWs depend on the relative strength of the different interparticle forces, which are due to normal, shear, and bending rigidities at the contacts. Interesting features of these SAWs are revealed, such as possible existence of the zero-group-velocity (ZGV) SAWs. The nature and discrete displacement profiles of the SAWs are described as a function of the parameters controlling the dispersion curves. In particular, the importance of bending rigidity is demonstrated, as the evolution of ZGV SAWs as well as the existence of SH-type SAWs strongly depend on bending interaction between beads. These analytical descriptions of SAWs dependence on the contact parameters, yield a possible comparison of the derived theoretical predictions with those of the simplest case of reduced Cosserat theory. The obtained results confirm that all effects of the material inhomogeneity are not correctly modeled in Cosserat theory because the spatial scale of the inhomogeneity, and consequently multiple scattering of the waves, are not accounted for. \\

In Section~\ref{ch4:sectRayleigh} Rayleigh-type waves are studied. SH-type waves are analyzed in Section~\ref{ch4:sectSH}. In both cases, the theoretical analysis shows the existence of SAWs propagating at the (010) surface along [100] direction and at the (110) surface along [1$\bar{1}$0] direction. Finally, comparison of SAWs in the granular crystals with those known predicted by the reduced Cosserat theory, is performed in Section~\ref{ch4:sectCosserat}. Some limitations of the Cosserat theory are then highlighted. 

\section{Rayleigh-type surface waves} \label{ch4:sectRayleigh}
\subsection{Dispersion curves of the propagating modes}
\begin{figure}[!ht]  
\centering
\includegraphics[scale=0.34]{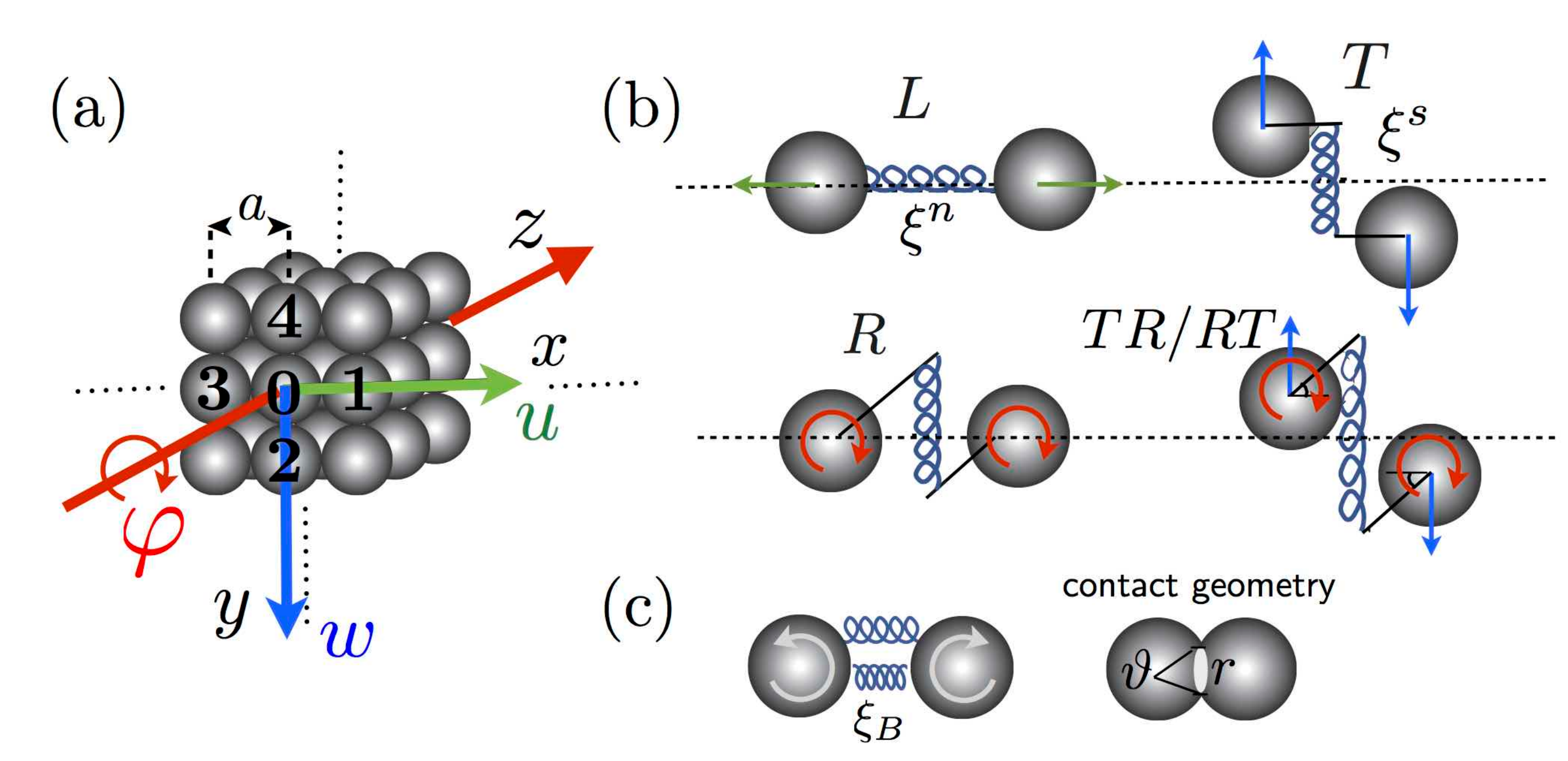}  
\caption{(Color online) (a) Schematic representation of the granular phononic crystal. $u$ denotes the longitudinal displacement along the $x$-axis, $w$ denotes the translational displacement motion along the $y$-axis and $\varphi$ denotes the rotational motion around the $z$-axis. (b) Illustration of the various possible motions of the beads, which are activating different contact springs/rigidities contributing to normal, shear and bending type interactions. (c) Schematic representation of the bending rigidity.}
\label{figC4:structure}
\end{figure}

The granular phononic crystal is made of spherical particles distributed periodically on a cubic lattice with periodicity $a$. The radius, mass and momentum of inertia of the particles are denoted by $R_c$, $m$ and $I$, respectively. The particles possess two translational and one rotational \textit{dofs}, Fig.~\ref{figC4:structure}. For the analysis of the plane Rayleigh-type SAWs, which are 2D motions of the crystal, the considered crystal is equivalent to the 2D one studied in~\cite{Pichard12}. Normal and shear forces at the contacts between two adjacent particles are described with springs of constant rigidity $\xi^n$ and  $\xi^s$, respectively. The elongations of the springs introduce forces and momenta that induce the motion of the particles: the displacements $u$ along the $x$-axis, $w$ along the $y$-axis and the rotation $\varphi$ around the $z$-axis. Different possible motions of two neighboring particles are illustrated in Fig.~\ref{figC4:structure}(b). The transversal, longitudinal, rotational and combined transversal/rotational motions are denoted by $T$, $L$, $R$, and $T/R$, respectively. Two spatially distinct normal springs of rigidity $\xi^B$ are introduced in Fig.~\ref{figC4:structure}(c) to model the effect of bending rigidity at the contact, i.e., of the interaction opposing the rotation of the two contacting beads in opposite directions~\cite{Tournat11}.\\
The complete derivation of the bulk dispersion relations for bulk acoustic waves can be found in~\cite{Pichard12}. The substitution of the plane wave solutions into the equations of motion leads to the eigenvalue problem
\begin{equation} \label{eqC4:detS}
\mathbf{S} \mathbf{v} = 0 \ ,
\end{equation}
where $\mathbf{v} =\begin{pmatrix} A_u \\ A_w \\ A_{\Phi} \end{pmatrix}$ is the amplitude vector, with $\Phi=R_c\varphi$ and $\mathbf{S}$ is the dynamical matrix defined by

\begin{equation} \label{eqC4:matrixS}
\mathbf{S}=
\left(\begin{array}{ccc}
\scriptstyle - \eta \sin^2 q_x - \sin^2q_y + \Omega^2 & \scriptstyle 0 &\scriptstyle \text{j} \sin q_y \cos q_y  \\ \scriptstyle 0 & \scriptstyle - \eta \sin^2q_y - \sin^2 q_x+ \Omega^2  &\scriptstyle -\text{j} \sin q_x \cos q_x \\ \scriptstyle-\text{j} p \sin q_y \cos q_y &\scriptstyle \text{j} p \sin q_x \cos q_x &\scriptstyle -p(\cos ^2q_x + \cos ^2q_y)-4p_B p (\sin^2 q_x + \sin^2 q_y)+ \Omega^2 
\end{array}\right) \ ,
\end{equation}

wherein $\Omega = \omega / \omega _0$ is the reduced frequency with $\omega _0 = 2 \sqrt{\xi^s /m}$, j is the imaginary unit, $q_{x,y}=k_{x,y} a /2$ are the normalized wave numbers, $p=mR_c^2 / I$, $\eta = \xi ^n / \xi^s$, $p_B=\dfrac{\vartheta^2 }{2}\dfrac{ \xi^B}{\xi^s}$ and $\vartheta$ is the angular contact dimension (see Fig.~\ref{figC4:structure}(c)). Note that, from classical mechanics $mR_c^2\ge 1$ and, as consequence, $p\ge 1$. In the case of a homogeneously filled and empty sphere where all the mass is at the sphere periphery, $p$ is equal to $2.5$ and $1.5$, respectively. \\\\ 
At any point $x$ and $y$ in the crystal, the displacement and rotation components of the modes are assumed to be in the form
\begin{equation} \label{ch4:eq_disp}
\begin{pmatrix}
u \\ w \\ \Phi
\end{pmatrix}
_{l,n}
=
\begin{pmatrix}
A_u \\ A_w \\ A_{\Phi}
\end{pmatrix}
e^{\text{j} \omega t -2\text{j} q_x\,l -  2\text{j} q_y\,n}
=A_{\Phi}
\begin{pmatrix}
 \alpha \\ \beta \\ 1 
 \end{pmatrix}
e^{\text{j} \omega t -2\text{j} q_x\,l -  2\text{j} q_y\,n} \ ,
\end{equation}
where $\alpha=\dfrac{\text{j}\sin q_y \cos q_y}{\eta \sin^2 q_x+\sin^2 q_y-\Omega^2} $ is the ratio between the longitudinal $A_u$ and rotational $A_{\Phi}$ amplitudes, $\beta=-\dfrac{\text{j}\sin q_x \cos q_x}{\eta \sin^2 q_y+\sin^2 q_x-\Omega^2}$ is the ratio between the transversal $A_w$ and rotational $A_{\Phi}$ amplitudes and $l$, $n$ refers to the particle position along $x$-axis and $y$-axis, respectively, measured in integer numbers of interparticle distances. \\\\
 Nontrivial solutions of Eq.~(\ref{eqC4:detS}) require that
\begin{equation} \label{C4eq:det1}
|S_{j,i}|=0 \ , \quad \text{with} \quad j,i=1,2,3 \ .
\end{equation}
For a given set of parameters $p$, $\eta$, $p_B$ and wave number $q_x$, Eq.~(\ref{C4eq:det1}) constitutes a relationship between the frequency $\Omega$ and the wave number $q_y$, which can be written as a cubic equation for $Y=\sin^2 q_y$ and for $\Omega^2$, see development in appendix \ref{app:detS}. Since it is a cubic equation in $\sin^2 q_y$ and in $\Omega^2$, for a given frequency $\Omega$ correspond six wave numbers $q_y$. Each pair of wave numbers describes either two waves propagating in opposite directions or two evanescent waves with opposite directions of the amplitude decay.  
\begin{figure}[!ht]  
\centering
\includegraphics[scale=0.47]{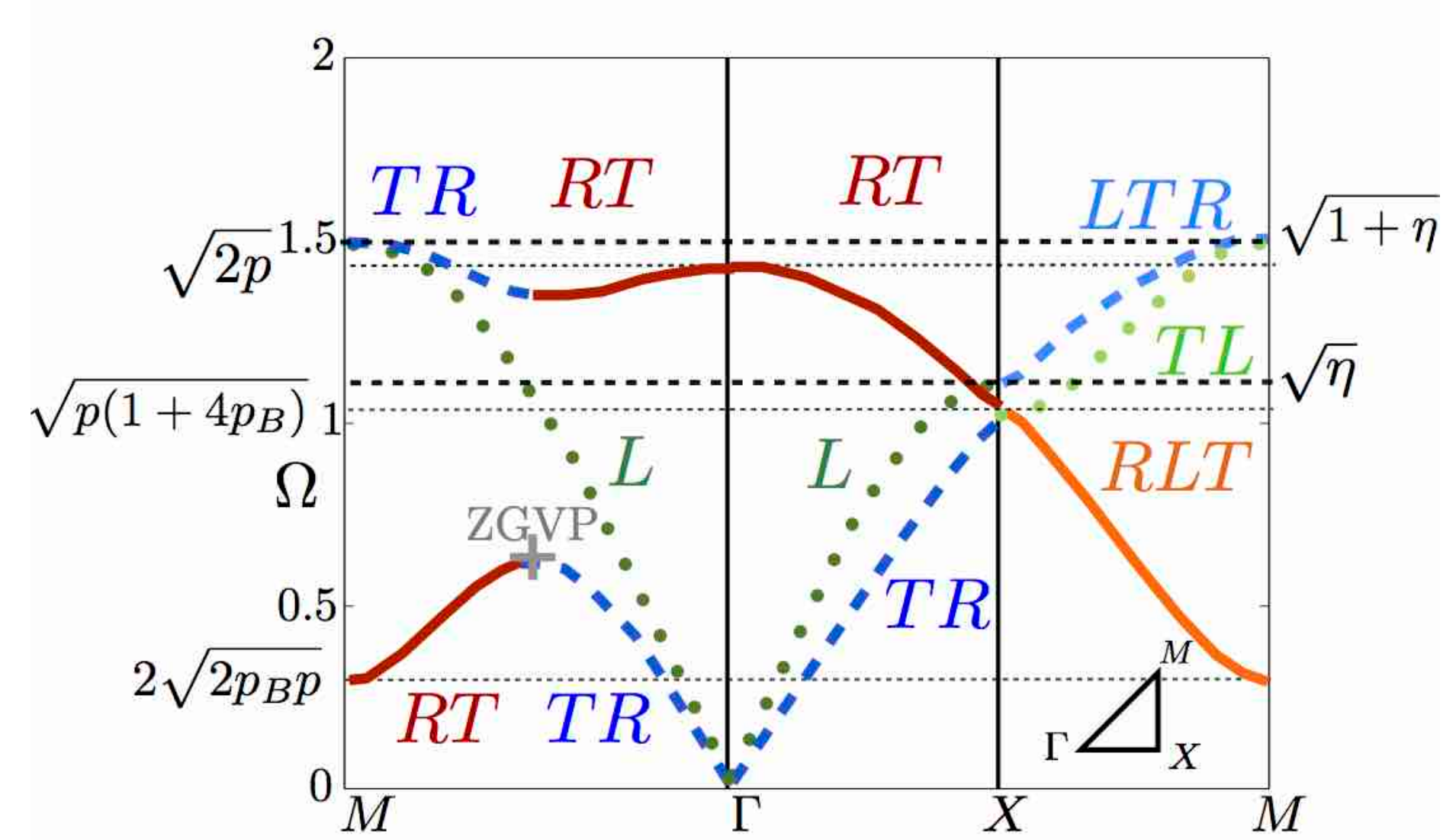}  
\caption{(Color online) Dispersion curves for the plane bulk waves possessing two translational and one rotational \textit{dofs}, obtained for $\eta=1.2$, $p=1$ and $p_B=0.01$. Solid curves correspond to coupled displacement-rotation modes (with a predominance of rotation), dashed curves correspond to coupled displacement-rotation modes (with a predominance of displacement), and dotted curves correspond to pure displacement modes.}
\label{figC4:dispVolume}
\end{figure}
Figure~\ref{figC4:dispVolume} presents the dispersion curves of the propagating modes obtained with $\eta=1.2$, $p=1$ and $p_B=0.01$. Each of the eigenmodes of the granular phononic crystal motion consists of three components, the longitudinal motion $L$, the transversal motion $T$, and the rotational motion $R$. The plotted eigenvalues have been colored accordingly to the eigenvectors that have been classified, and the nature of the modes is labeled. The continuous red-orange lines correspond to coupled displacement-rotation modes with a predominance of rotation ($RT$, $RLT$), the dashed blue lines correspond to coupled displacement-rotation modes with a predominance of displacement ($TR$, $LTR$), and the dotted green lines correspond to pure displacement modes ($L$, $TL$). Ref.~\cite{Pichard12} provides a complete description of the dispersion curves as a function of the parameters $p$, $\eta$ and $p_B$. A remarkable feature of the lowest transversal/rotational mode along the $\Gamma M$ direction is the existence of two group velocity regions separated by a zero-group velocity point (ZGVP, see Fig.~\ref{figC4:dispVolume}), resulting in birefraction phenomenon. The position and existence of this ZGVP strongly depends on parameters $p$, $\eta$, and $p_B$. The developed theory explains the physical origin of these modes, which are due to interaction/repulsion of the transversal and rotational motions, leading to hybridized rotational/transversal modes. The description of the ZGVP and of these nonmonotonous modes as well as their dependence on the bending rigidity parameter $p_B$, is presented in details in~\cite{Pichard12}.

\subsection{Boundary conditions for Rayleigh-type SAWs propagating at the (010) free surface along [100] direction} \label{ch4:sectBC}
\begin{figure}[!ht]  
\centering
\includegraphics[scale=0.34]{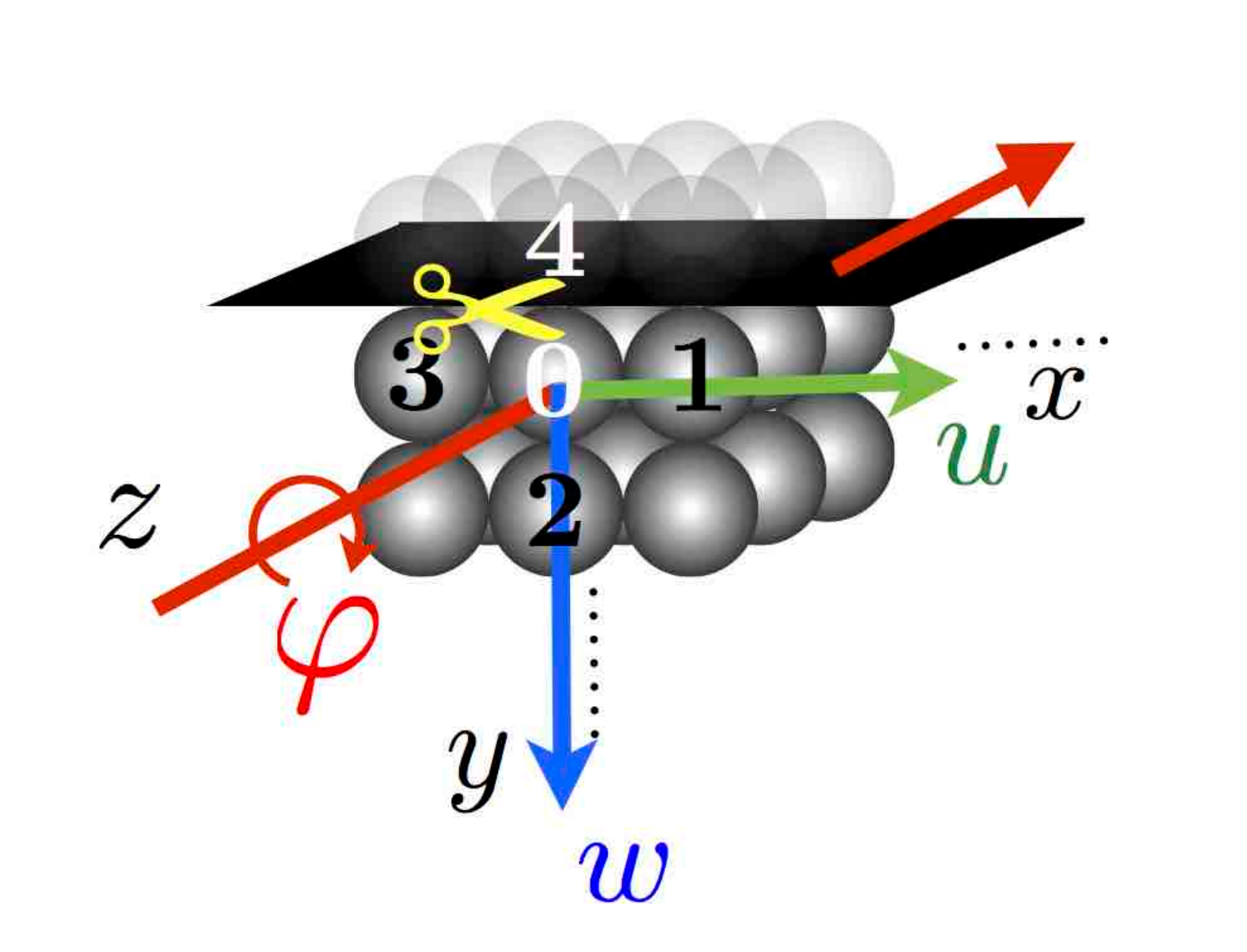}  
\caption{(Color online) Schematic representation of the (010) surface and the propagation direction ($x$-direction) for the Rayleigh-type SAWs.}
\label{figC4:surface}
\end{figure}
In this section we study the Rayleigh-type SAWs at the mechanically free surface of the granular crystal, which is normal to the $y$-axis, i.e., on (010) surface. Waves propagating along $x$-axis, i.e., in the [100] direction, are sought. The boundary conditions are derived from the removal of all particles on one side of the boundary layer. In this configuration, the surface modes, whose amplitudes decrease away from the boundary, i.e., along the $y$-axis, have an attenuation defined by the imaginary part of the wave number $q_y$. Mechanically free boundary conditions are the absence of forces and of momentum between $0$-th and $4$-th beads, Fig.~\ref{figC4:surface}. These conditions are satisfied if the contact between the $0$-th and $4$-th beads is not strained, i.e., the normal and shear springs are not deformed/elongated while the beads $1$ and $4$ are rotating in opposite directions, preventing activation of the bending rigidity of the contact. Thus the mechanically free boundary conditions can be formulated mathematically as follows\\
\begin{itemize}
\item No longitudinal spring elongation,
\begin{equation}\label{eqC4:BC_GX_1}
w_0-w_4=0 \ . 
\end{equation}

\item No shear spring elongation,
\begin{equation}\label{eqC4:BC_GX_2}
u_4-u_0 +(\Phi _4 + \Phi _0)=0 \ .
 \end{equation}

\item No rotation, which can activate bending rigidity,
\begin{equation}\label{eqC4:BC_GX_3}
\Phi _4 - \Phi _0=0 \ .
\end{equation}
\end{itemize}
The boundary conditions should be satisfied by the bulk modes whose amplitudes decrease as $n$ increases. Therefore, the localization of surface waves should operated by complex wave numbers with a negative imaginary part, i.e., by three of the six wave numbers given by Eq.~(\ref{C4eq:det1}). When the solutions for the wave number $q_y$ are purely real, the waves are propagative. Three evanescent modes, whose frequency lies in the forbidden band for propagating waves, and characterized by a complex-valued wave number in such a way that the amplitude of the mode decays with increasing $y$-coordinate, should be coupled to satisfy the derived boundary conditions.\\
\\
If $A_{\Phi_{i}}$ are the amplitudes of $\Phi$ in the 3 modes, the displacement and rotation components of the modes can be written in the following form
\begin{subequations}\label{eqC4:ampl}
\begin{gather}
u_{l,n}=\sum\limits_{i=1}^{3} A_{\Phi_{i}} \, \alpha_i \, e^{\text{j}\omega t} e^{-2\text{j}l q_x}e^{-2\text{j}n q_{y_{i}}} \ ,\\
w_{l,n}=\sum\limits_{i=1}^{3} A_{\Phi_{i}} \, \beta_i \, e^{\text{j}\omega t} e^{-2\text{j}l q_x}e^{-2\text{j}n q_{y_{i}}} \ ,\\
\Phi_{l,n}=\sum\limits_{i=1}^{3} A_{\Phi_{i}}  \, e^{\text{j}\omega t} e^{-2\text{j}l q_x}e^{-2\text{j}n q_{y_{i}}} \ ,
\end{gather}
\end{subequations}
with $\alpha_i=\dfrac{\text{j}\sin q_{y_i} \cos q_{y_i}}{\eta \sin^2 q_x+\sin^2 q_{y_i}-\Omega^2} $ and $\beta_i=-\dfrac{\text{j}\sin q_x \cos q_x}{\eta \sin^2 q_{y_i}+\sin^2 q_x-\Omega^2}$, with $i=1,2,3$ for the first, second and third mode, respectively, while $q_x$ has the physical meaning of a surface wave number.\\\\
The substitution of these amplitudes, Eqs.~(\ref{eqC4:ampl}), into the boundary conditions, Eqs.~(\ref{eqC4:BC_GX_1}) -~(\ref{eqC4:BC_GX_3}), leads to
\begin{subequations}
\begin{gather}
\sum\limits_{i=1}^{3} A_{\Phi_{i}} \, \beta_i(1-e^{2\text{j}q_{y_i}})=0 \ ,\\
\sum\limits_{i=1}^{3}\left[ A_{\Phi_{i}} \, \alpha_i(e^{2\text{j}q_{y_i}}-1)+A_{\Phi_{i}}(1+e^{2\text{j}q_{y_i}})\right]=0 \ ,\\
\sum\limits_{i=1}^{3} A_{\Phi_{i}} \, (e^{2\text{j}q_{y_i}}-1)=0 \ ,
\end{gather}
\end{subequations}
which can be rewritten in the following form,
\begin{equation}\label{eqC4:probS2}
\mathbf{S_2} \mathbf{v_2} =0 \ ,
\end{equation}
with $\mathbf{v_2} =\begin{pmatrix}A_{\Phi_{1}} \\ A_{\Phi_{2}} \\ A_{\Phi_{3}} \end{pmatrix}$ and
\begin{equation} \label{matrixS2}
\mathbf{S_2}=
\left(\begin{array}{ccc}
\beta_1(1-e^{2\text{j}q_{y_1}}) & \beta_2(1-e^{2\text{j}q_{y_2}}) & \beta_3(1-e^{2\text{j}q_{y_3}}) \\
 \alpha_1(e^{2\text{j}q_{y_1}}-1)+1+e^{2\text{j}q_{y_1}} & \alpha_2(e^{2\text{j}q_{y_2}}-1)+1+e^{2\text{j}q_{y_2}} & \alpha_3(e^{2\text{j}q_{y_3}}-1)+1+e^{2\text{j}q_{y_3}} \\
e^{2\text{j}q_{y_1}}-1 & e^{2\text{j}q_{y_2 }}-1 & e^{2\text{j}q_{y_3}}-1
 \end{array}\right) \ .
\end{equation}
In order to have nontrivial solutions of Eq.~(\ref{eqC4:probS2}), the following equation must be satisfied
\begin{equation} \label{eqC4:det2}
|S_{2_{j,i}}|=0 \quad j,i=1,2,3 \ .
\end{equation}
For a set of parameters $p$, $p_B$, $\eta$ and for a propagation wave number specified by $q_x$, the solutions $\Omega$ and the corresponding $q_{y_i}$ of the surface modes are obtained from the simultaneous solutions of Eqs.~(\ref{C4eq:det1}) and~(\ref{eqC4:det2}). These surface modes are discussed in the following section. \\\\
According to Eqs.~(\ref{eqC4:ampl}), the amplitudes of the longitudinal $u_{l,n}$, transversal $w_{l,n}$ and rotational $\Phi_{l,n}$ discrete displacements of the modes as a function of the particle position $(l,n)$ in the crystal can be determined by combining the bulk modes with projections of the wave vector along the $y$-axis, $q_{y_1}$, $q_{y_2}$ and $q_{y_3}$
{\small\begin{equation}
\begin{pmatrix}
u_i \\ w_i \\ \Phi_i
\end{pmatrix}
_{l,n}
=
 A_{\Phi}\left[ Z_1 \begin{pmatrix} \alpha_1 \\ \beta_1 \\ 1 \end{pmatrix} e^{\text{j}\omega t} e^{-2\text{j}l q_x}e^{-2\text{j}n q_{y_{1}}} + Z_2 \begin{pmatrix} \alpha_2 \\ \beta_2 \\ 1 \end{pmatrix} e^{\text{j}\omega t} e^{-2\text{j}l q_x}e^{-2\text{j}n q_{y_{2}}} + \begin{pmatrix} \alpha_3 \\ \beta_3 \\ 1 \end{pmatrix} e^{\text{j}\omega t} e^{-2\text{j}l q_x}e^{-2\text{j}n q_{y_{3}}} \right] \ ,
\end{equation}}
with $Z_1=\dfrac{A_{u}}{ A_{\Phi}}=-1-Z_2$ and $Z_2=\dfrac{A_{w}}{ A_{\Phi}}=\dfrac{(\beta_1-\beta_3)(1-e^{2\text{j}q_{y_3}})}{(\beta_1-\beta_2)(e^{2\text{j}q_{y_2}}-1)}$.\\

\begin{figure}[!ht]
\centering
\includegraphics[scale=0.4]{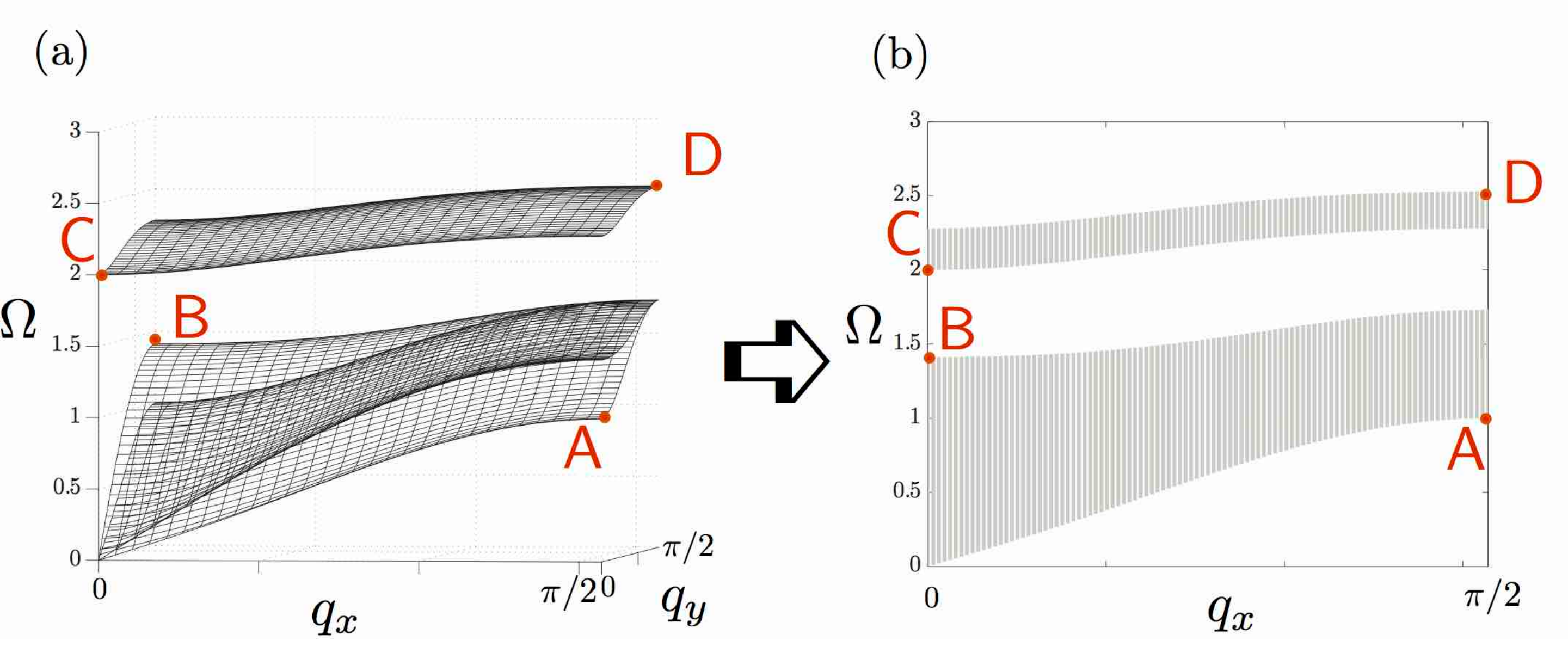}  
\caption{(Color online) (a) 3D dispersion curves of the crystal for $\eta=2$, $p=2$ and $p_B=0.4$. (b) Projected bulk bands along [100] direction, i.e., $x$ direction.}
\label{fig:projected}
\end{figure}

The domain of the admissible wave numbers and frequencies where SAWs could be sought, through the solution of Eqs.~(\ref{C4eq:det1}) and~(\ref{eqC4:det2}), could be importantly reduced by presenting the dispersion curves of the bulk modes in the granular crystal in projected band diagram~\cite{Joannopoulos95}.
As illustrated in Fig.~\ref{fig:projected}, to construct a band diagram projected onto the $q_x$ direction, i.e., on the direction of SAWs propagation, the value of $q_x$ is fixed and the frequencies corresponding to all possible real projections of the wave number $q_y$, i.e., of the bulk modes on the $y$-axis, are plotted in the same graph. For example, the obtained projected band diagram in the case of $p=2$, $\eta=2$ and $p_B=0.4$ is represented in Fig.~\ref{fig:projected}(b). In this band diagram, gray shaded regions define allowed/propagating phononic bands, while empty regions define band gaps. Note that the reprensentation along the $\Gamma X M \Gamma$ directions involves only the solutions for bulk modes on the edge of an irreductible Brillouin zone (1D calculation), while the band diagram of Fig.~\ref{fig:projected}(b) requires computation of all the solutions inside an irreductible Brillouin zone (2D calculation). Although the $\Gamma X M \Gamma$ representation gives consistent definitions of complete band gaps the projected band diagram in Fig.~\ref{fig:projected}(b) is preferred when detailed information is required. To determine the regions allowed for surface waves, the projected bands diagram along the $q_x$ direction are chosen in the following analyses. In fact, it is necessary to use the projected diagram for the analysis of possible surface waves, because the surface waves cannot lie in a propagative band and should be located between them. Otherwise SAWs  emit bulk modes and are evanescent, i.e., decay along their propagation path, $x$-axis.

\subsection{Pure longitudinal mode}
\begin{figure}[!ht]
\centering
\includegraphics[scale=0.36]{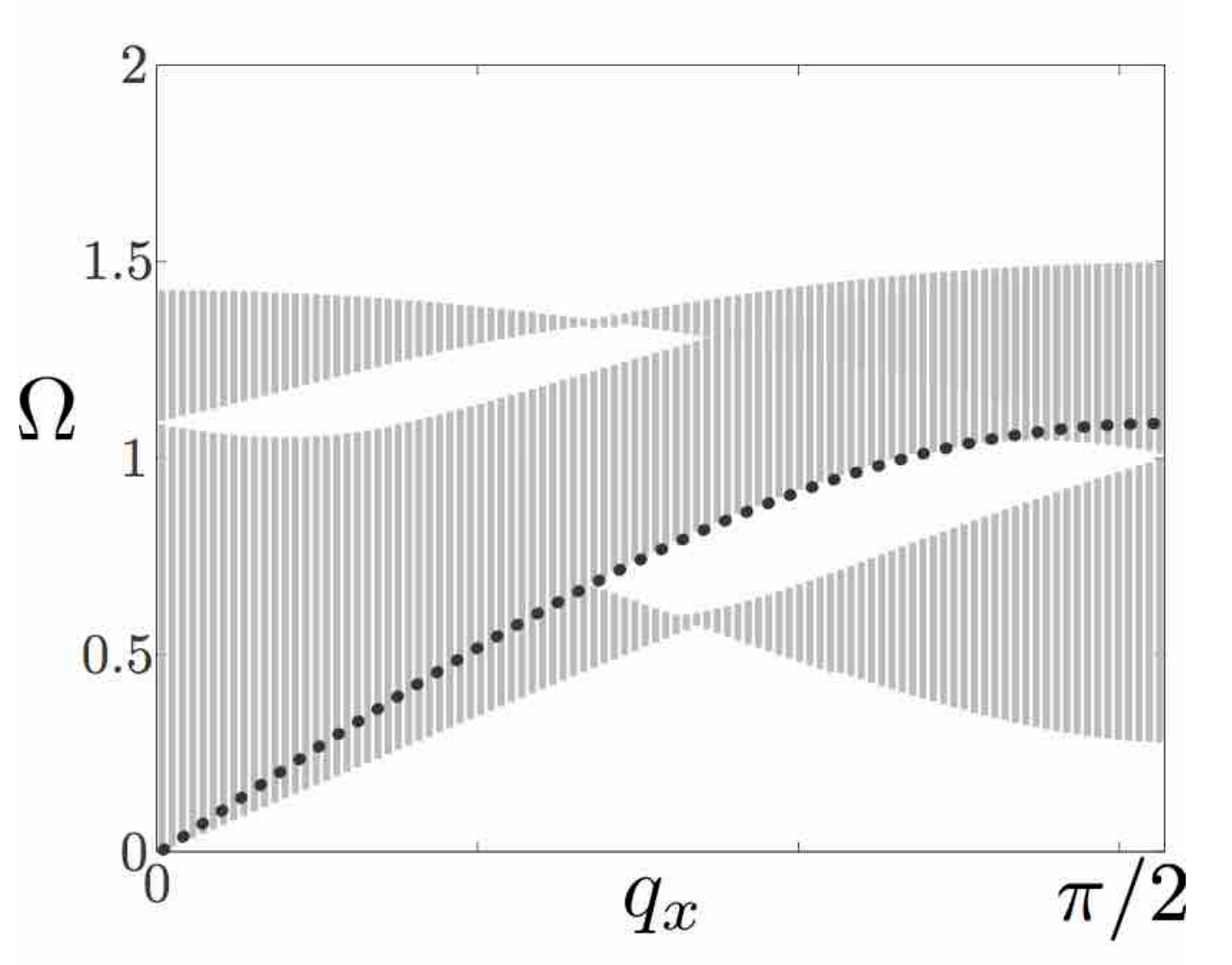}  
\caption{(Color online) Dispersion curves along the $q_x$ direction of the crystal for $\eta=1.2$, $p=1$ and $p_B=0.01$. The shaded areas represent the projected bulk bands along [100], i.e., $x$ direction. The dotted black curve represents the mode $\Omega^2=\eta \sin^2 q_x$.}
\label{fig:modeplan}
\end{figure}
From the development of the boundary conditions determinant Eq.~(\ref{eqC4:det2}), it follows that a pure longitudinal mode $\Omega^2=\eta \sin^2 q_x$ propagating along the $x$-axis satisfies the boundary conditions at the considered mechanically free surface of the cubic crystal (the development can be found in appendix~\ref{App:planemode}). The dispersion curve of this mode is shown in black dotted curve in Fig.~\ref{fig:modeplan}. This is a pure longitudinal mode skimming along the surface ($A_\Phi=A_w=0$), which exhibits the same dispersion than the pure longitudinal mode propagating along the $\Gamma X$ direction of the crystal~\cite{Pichard12}. This mode is not coupled with the rotational and transverse waves because the same relative motion of the neighbor particles along the $x$-axis at all distances from the surface does not lead to deformation of shear springs. Physically, this corresponds to a wave propagating in a material with a zero Poisson coefficient, showing no expansion/contraction in the direction orthogonal to the axis of its compression. 

\subsection{Surface modes description}
\begin{figure}[!ht]
\centering
\includegraphics[scale=0.35]{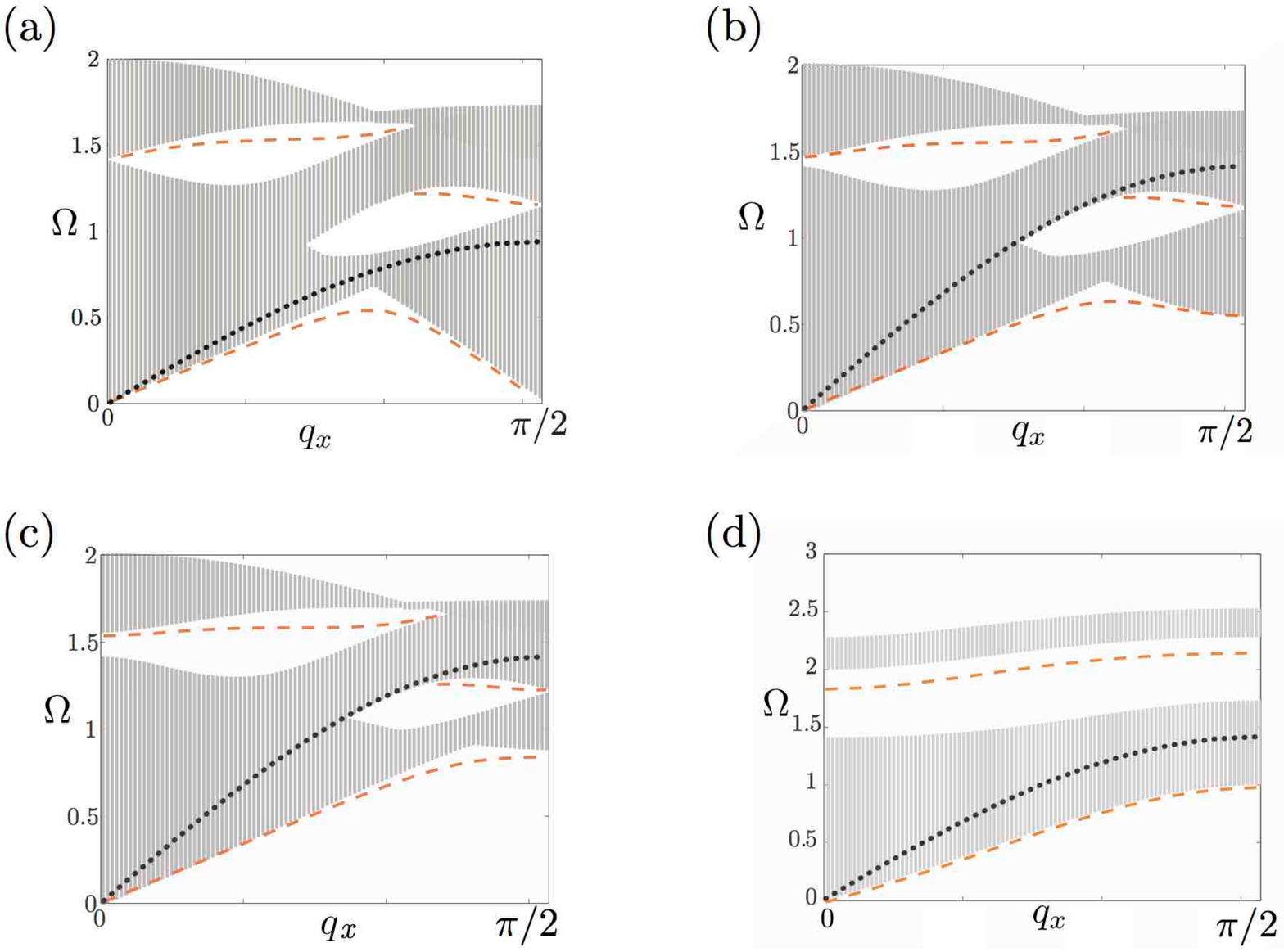}  
\caption{(Color online) Dispersion curves along the $q_x$ direction of the crystal for $\eta=2$, $p=2$ and (a) $p_B=0$, (b) $p_B=0.02$, (c) $p_B=0.05$ and (d) $p_B=0.4$. The shaded areas represent the projected bulk bands along [100], i.e., $x$ direction. The dashed orange curves represent the surface modes and the dotted black curve represents the longitudinal bulk mode propagating along the surface.}
\label{fig:evolutionpB}
\end{figure}

\begin{figure}[!ht]
\centering
\includegraphics[scale=0.38]{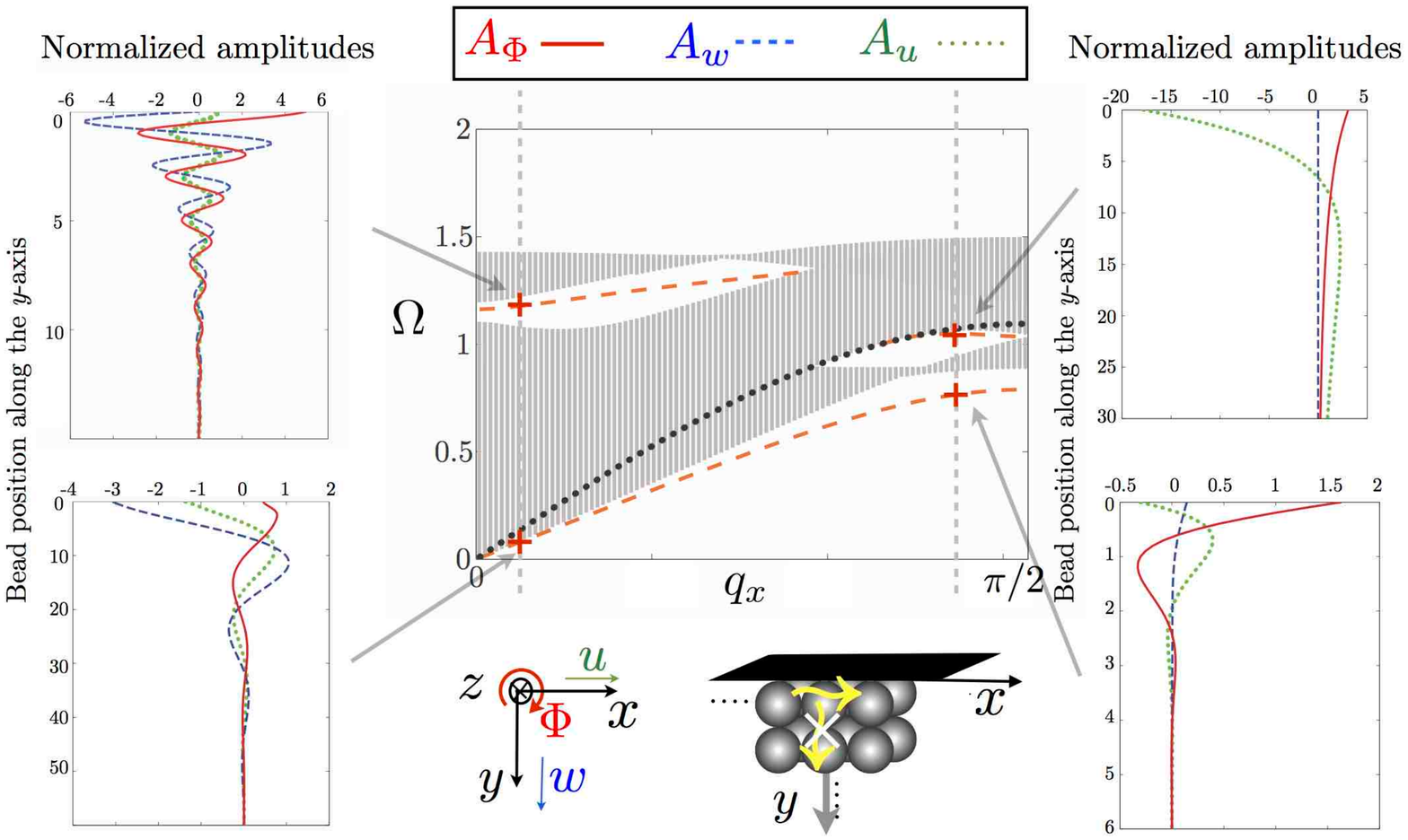}  
\caption{(Color online) Discrete displacement and rotation profiles along the $y$-axis, i.e., perpendicular to the surface, for two distinct surface modes correponding to each of the two different wave numbers $q_x$.}
\label{fig:amplqy}
\end{figure}

Surface waves are calculated for fixed sets of parameters $q_x$, $p$, $p_B$ and $\eta$ by simultaneous solutions of Eqs.~(\ref{C4eq:det1}) and Eqs.~(\ref{eqC4:det2}). Fig.~\ref{fig:evolutionpB} presents the evolution of the obtained surface modes for $p=\eta=2$ by increasing the bending rigidity parameter $p_B$.  Example of discrete displacement profiles of the surface modes along the $y$-axis are given in Fig.~\ref{fig:amplqy}, for two fixed wave numbers $q_x$. The projected bulk bands along [100], i.e., $x$ direction, are represented by shaded areas. The surface modes are represented in dashed orange curves, and the pure longitudinal mode $\Omega^2=\eta \sin^2 q_x$ propagating along the surface is drawn in dotted black curve. 

One surface mode propagates below the first propagating band, while, one surface or two surface modes can exist between the upper propagating bands of bulk modes depending on the parameters values. In absence of bending rigidity, Fig.~\ref{fig:evolutionpB}(a), the mode at low frequencies has a nonmonotonous behavior. It presents a ZGVP and vanishes at $q_x = \pi/2$. At the ZGVP the energy that could be pumped into this mode does not propagate away from the region of excitation on the surface. Since at a ZGVP the phase velocity and so the wavelength remain finite, while the group velocity is zero, the energy can be locally trapped in the source area without any transfer to the adjacent medium. The ZGVP was investigated earlier in the case of Lamb waves~\cite{Prada05}, i.e., in finite thickness structures. Here, this type of singularity is theoretically predicted for semi-infinite medium, i.e., in the case of Rayleigh-type surface waves. When increasing the bending rigidity parameter $p_B$, the frequency value of this mode increases at $q_x=\pi/2$, Fig.~\ref{fig:evolutionpB}(b)(c), and then, for sufficiently large value of $p_B$, the ZGVP disappears, Fig.~\ref{fig:evolutionpB}(d). The physical origin of the predicted zero-group-velocity surface acoustic wave is in the existence of bulk modes where coupling/hybridization of rotational and transverse motion of the beads produces non-monotonous $TR$/$RT$ modes with ZGVP, Fig.~\ref{figC4:dispVolume}. In classical surface Rayleigh waves in the isotropic solids the bulk longitudinal and transverse waves, constituting the SAWs, are coupled by the mechanically free surface. On the mechanically free surface of the granular crystal the longitudinal wave is coupled to $TR$/$RT$ mode and the non-monotonous character of the latter can be transformed in the non-monotonous dispersion relation for the Rayleigh-type SAW and the existence of the ZGV SAWs. It could be expected that ZGVPs for SAWs, similarly to the ZGVPs in Lamb modes, could find application in nondestructive testing of the materials~\cite{Blitz96}.

\subsection{Rayleigh-type SAWs propagating at the (110) surface along [1$\bar{1}$0] direction}
\begin{figure}[!ht]
\centering
\includegraphics[scale=0.4]{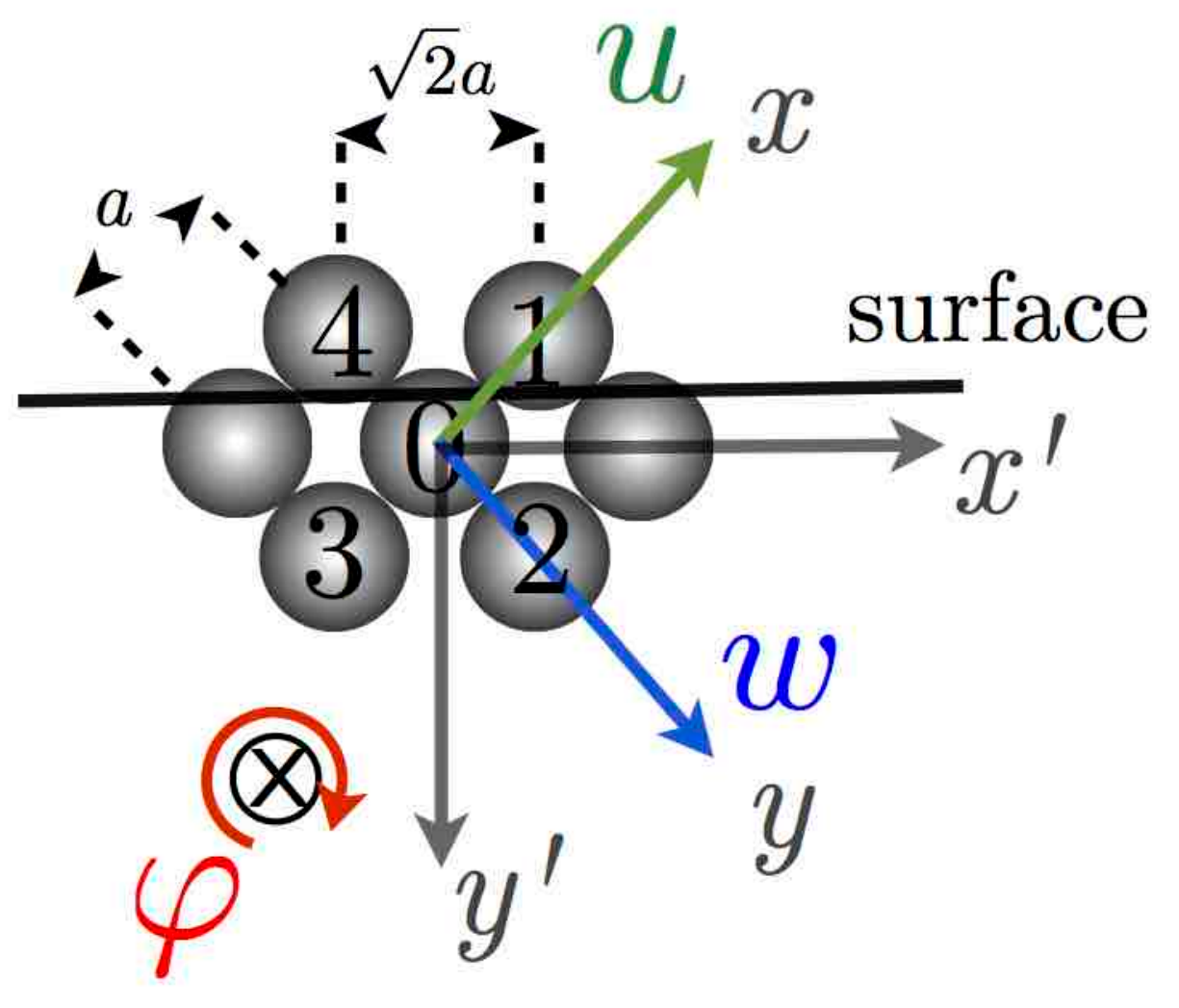}
\caption{(Color online) Schematic representation of the surface.}
\label{figC4:diag}
\end{figure}
\begin{figure}[!ht]
\centering
\includegraphics[scale=0.37]{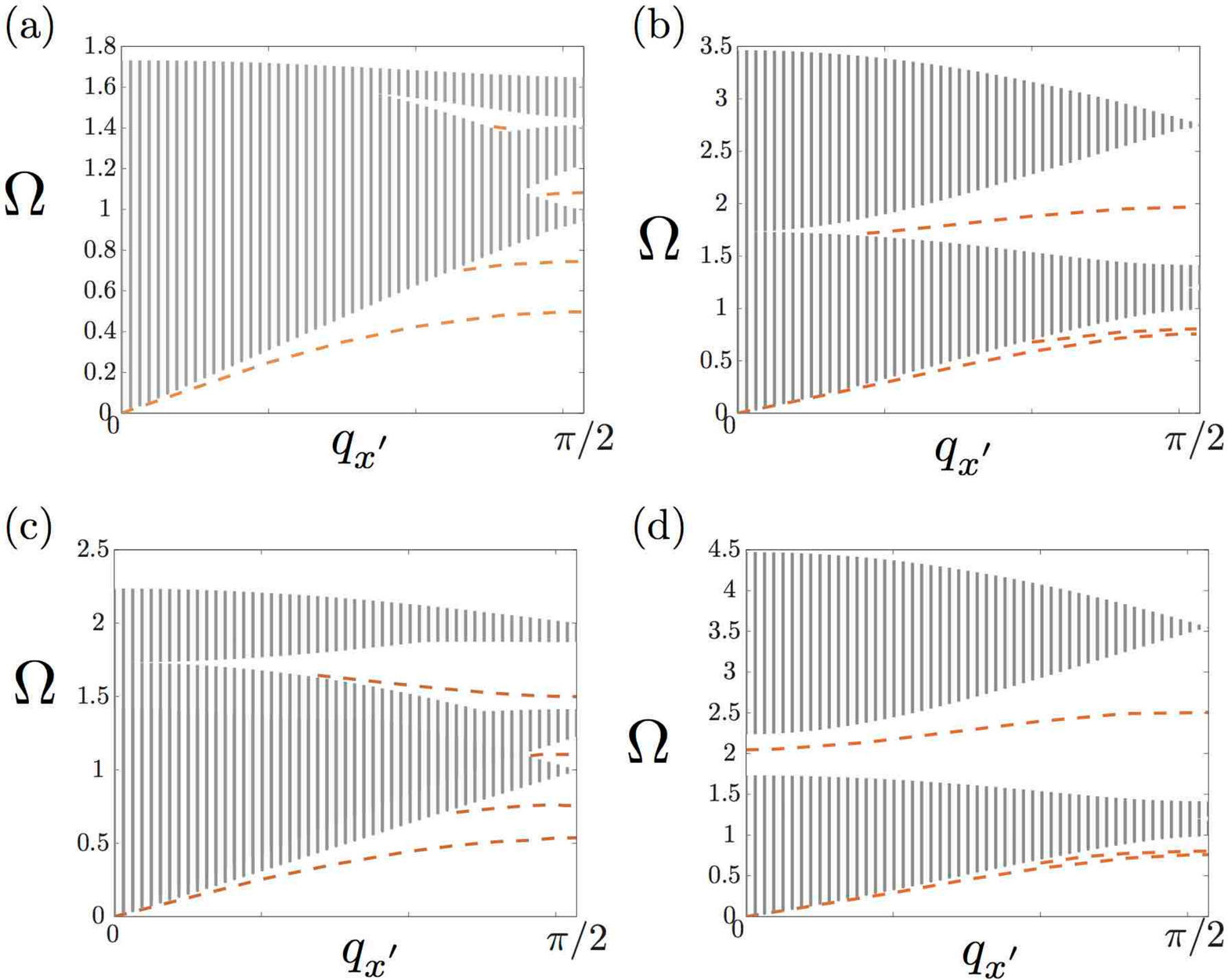}
\caption{(Color online) Dispersion curves along the diagonal direction of the crystal for $\eta=2$ in the case of empty and filled spheres, (a) $p=1.5$ and $p_B=0.1$, (b) $p=1.5$ and $p_B=1$, (c) $p=2.5$ and $p_B=0.1$ and (d) $p=2.5$ and $p_B=1$. The shaded areas represent the projected bulk bands along [1$\bar{1}$0] and the dashed orange curves represent the surface modes.}
\label{figC4:diag_p15_eta2}
\end{figure}
Rayleigh-type SAWs propagating at the (110) surface along [1$\bar{1}$0] direction are analyzed in this part. Figure~\ref{figC4:diag} illustrates the position of the surface, which is parallel to the diagonal of the cubic crystal, and introduces new axes $x^{'}$ and $y^{'}$.  The complete derivation of the bulk dispersion is not given in details here, but the reasoning is the same as in Sect.~\ref{ch4:sectBC}. In this case, the eigenvalue problem arising from the substitution of the plane wave solutions into the equations of motion is
\begin{equation} \label{eqC4:detS_diag}
\mathbf{S_{\textrm{diag}}} \mathbf{v} = 0 \ ,
\end{equation}
with $\mathbf{v} =\begin{pmatrix} A_u \\ A_w \\ A_{\Phi} \end{pmatrix}$ and
\begin{equation} \label{eqC4:matrixS_diag}
\mathbf{S_{\textrm{diag}}}=
\left(\begin{array}{ccc}
\scriptstyle -\frac{1}{2}(\eta+1)(1-\cos q_x^{'} \cos q_y^{'} )+\Omega^2 & \scriptstyle -\frac{1}{2}(\eta-1) \sin q_x^{'} \sin q_y^{'} & \scriptstyle \frac{1}{\sqrt{2}}\textrm{j} \cos q_x^{'}\sin q_y^{'}  \\ \scriptstyle -\frac{1}{2}(\eta-1) \sin q_x^{'} \sin q_y^{'} &\scriptstyle -\frac{1}{2}(\eta+1)(1-\cos q_x^{'} \cos q_y^{'}) +\Omega^2  &\scriptstyle -\frac{1}{\sqrt{2}}\textrm{j} \sin q_x^{'} \cos q_y^{'} \\\scriptstyle -\frac{p}{\sqrt{2}}\textrm{j}\cos q_x^{'} \sin q_y^{'} & \scriptstyle \frac{p}{\sqrt{2}}\textrm{j} \sin q_x^{'} \cos q_y^{'} &\scriptstyle -(p+4p_B p)-(p-4p_Bp)\cos q_x^{'} \cos q_y^{'}+\Omega^2 
\end{array}\right) \ ,
\end{equation}
with $q_x^{'}$ and $q_y^{'}$ the normalized wave numbers along $x^{'}$ and $y^{'}$ axes, respectively.\\\\
Mechanically free boundary conditions are applied at the surface, i.e., the total forces of beads $1$ and $4$ acting on bead $0$ are zero. The amplitudes $A_{u_i}$, $A_{w_i}$ and $A_{\Phi_i}$ corresponding to one $q^{'}_{y_i}$ can be determined by
\begin{equation}
\dfrac{A_{u_i}}{\chi_i}=\dfrac{A_{w_i}}{\epsilon_i}=\dfrac{A_{\Phi_i}}{\zeta_i}=\Lambda_i,
\end{equation}
where $\chi_i$, $\epsilon_i$ and $\zeta_i$ are the cofactors of any row of the determinant of the dynamical matrix~(\ref{eqC4:matrixS_diag}) associated with $q^{'}_{y_i}$ ($i=1,2,3$) and where the $\Lambda_i$ are  determined from the boundary conditions. 
Hence, the general solution is
\begin{equation} \label{eq:exp_cofR}
\begin{pmatrix}
u\\  w\\ \Phi  
\end{pmatrix}
=
\sum\limits_{i=1}^{3} (\chi_i, \epsilon_i,  \zeta_i) \, \Lambda_i \, e^{\text{j} \omega t - \text{j} q^{'}_x \, x^{'} -\text{j}  q^{'}_y\,  y^{'}}.
\end{equation}
Substituting Eq.~(\ref{eq:exp_cofR}) into the boundary condition system leads to
\begin{equation}\label{eq:EVP2}
\sum\limits_{i=1}^{3} S_{2\textrm{diag}_{j,i}} \, \Lambda_i =0 \quad (i,j=1,2,3),
\end{equation}
where
\begin{equation}
\begin{split}
S_{2\textrm{diag}_{1,i}}&=\chi_i \dfrac{\eta+1}{\sqrt{2}}(1-\cos q_x^{'} e^{\textrm{j}q_{y_i}^{'}}) - \textrm{j}\epsilon_i \dfrac{\eta-1}{\sqrt{2}}\sin q_x^{'} e^{\textrm{j}q_{y_i}^{'}} -\zeta_i (1+\cos q_x^{'} e^{\textrm{j}q_{y_i}^{'}}),\\
S_{2\textrm{diag}_{2,i}}&= -\textrm{j}\chi_i \dfrac{\eta-1}{\sqrt{2}}\sin q_x^{'} e^{\textrm{j}q_{y_i}^{'}}+\epsilon_i\dfrac{\eta+1}{\sqrt{2}}(1-\cos q_x^{'} e^{\textrm{j}q_{y_i}^{'}})+\textrm{j} \zeta_i \sin q_x^{'} e^{\textrm{j}q_{y_i}^{'}},\\
S_{2\textrm{diag}_{3,i}}&=\chi_i\dfrac{1}{\sqrt{2}}(-1+\cos q_x^{'} e^{\textrm{j}q_{y_i}^{'}} )+\textrm{j}  \epsilon_i \dfrac{1}{\sqrt{2}} \sin q_x^{'} e^{\textrm{j}q_{y_i}^{'}}+\zeta_i\left[(-4p_B+1)\cos q_x^{'} e^{\textrm{j}q_{y_i}^{'}}  +1+4p_B\right].
\end{split}
\end{equation}

Surface waves can then be obtained by simultaneous fulfilment of Eqs.~(\ref{eqC4:detS_diag}) and~(\ref{eq:EVP2}).
Figure~\ref{figC4:diag_p15_eta2} presents the obtained dispersion curves for $\eta$=2 and by increasing bending rigidity in the case of a crystal made of empty ($p=1.5$) and filled ($p=2.5$) spheres. The shaded areas represent the projected bulk bands along [1$\bar{1}$0] and the dashed orange curves represent the surface modes. For all parameter values, two surface modes are found below the first propagative band and several branches lie in the gap between the upper propagative bands. Along this direction, the SAWs present a monotonous behavior.

\section{Shear-Horizontal (SH) type surface waves} \label{ch4:sectSH}
\subsection{Dispersion curves of the propagating modes}
\begin{figure}[!ht]  
\centering
\includegraphics[scale=0.34]{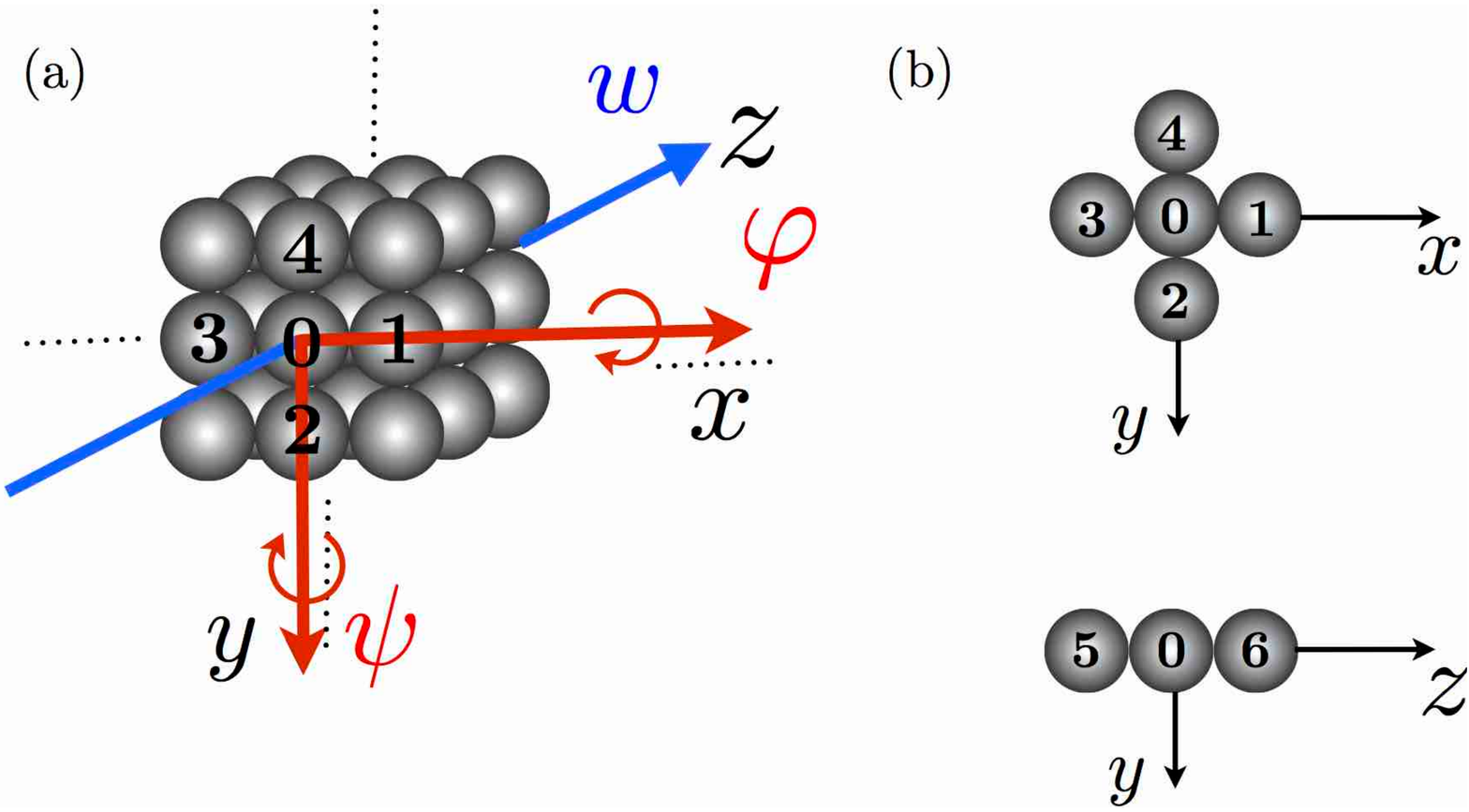}  
\caption{(Color online) (a) Schematic representation of the granular crystal. $w$ denotes the translational displacement motion along the $z$-axis and $\varphi$ (respectively $\psi$) the rotational motion around the $x$-axis (respectively $y$-axis). (b) Definition of the bead numbers along the different axes.}
\label{figC4:structure_SH}
\end{figure}
The studied granular phononic crystal is composed of spheres distributed periodically on a cubic lattice and possessing two rotational and one translational \textit{dofs}, Fig.~\ref{figC4:structure_SH}. The shear force at the contact between two adjacent particles is described by a spring of constant rigidity $\xi^s$. The elongation of the springs introduces forces and momenta that induce the motion of the particles: the rotation $\varphi$ around the $x$-axis, the rotation $\psi$ around the $y$-axis and the displacement $w$ along the $z$-axis.
The equations of motion of the central particle obtained by applying the Lagrange principle are given by
\begin{subequations} \label{eq_motionsphere}
\begin{gather}
m \ddot{w}_0 = - \xi^s \left[  \delta s_1+ \delta s_2+\delta s_3+ \delta s_4 \right] \ ,\\
I \ddot{\varphi}_0 = -\xi ^s R_c \left[ -\delta s_4 + \delta s_2 + \delta v_5 - \delta v_6 \right] + M_4 + M_2 \ ,\\
I \ddot{\psi}_0 = -\xi ^s R_c \left[ -\delta s_1 + \delta s_3 - \delta h_5 + \delta h_6 \right] + M_1 + M_3 \label{eq_motion_rot} \ ,
\end{gather}
\end{subequations}
where $m$ is the mass of the particle and $I$ is its momentum of inertia. The spring elongation in the transversal direction between the central and the ${\text{i}}$-th particle, i.e., the relative displacement between the $0$-th and the ${\text{i}}$-th particle at the contact point, is denoted by $\delta s_{\text{i}}$ and the momenta due to bending rigidity are denoted by $M_{\text{i}}$. The relative displacements are explicitly given by,
\begin{equation} \label{deltasphere}
\begin{split}
\delta s_1&=w_0-w_1 - R_c(\psi _0 + \psi _1) \ ,\\
\delta s_2&=w_0-w_2 + R_c(\varphi _0 + \varphi _2) \ ,\\
\delta s_3&=w_0-w_3 + R_c(\psi _0 + \psi _3) \ ,\\
\delta s_4&=w_0-w_4 - R_c(\varphi _0 + \varphi _4) \ .
\end{split}
\end{equation}

The springs oriented along the $x$-axis and $y$-axis for the contact between the $0$-th and $5$-th beads, and the $0$-th and $6$-th beads are active in shearing, Fig.~\ref{figC4:structure_SH}. The beads $5$ and $6$, oriented along the $z$-axis, rotate in the same direction and same angle as the central particle, i.e., the bending rigidity is not initiated because $\varphi_0=\varphi_5=\varphi_6$ and $\psi_0=\psi_5=\psi_6$. Because of the study of SH surface waves, which are a particular 2D motions of the crystal, there is no dependance on the $z$ coordinate, i.e., $w_0=w_5=w_6$. Then, the vertical $\delta v_{\text{i}}$ (along $y$-axis) and horizontal $\delta h_{\text{i}}$ (along $x$-axis) spring elongations between the central and the $5$-th and $6$-th beads are 
\begin{equation}\label{deltasphere2}
\begin{split}
\delta v_5&= 2R_c \varphi_0 \ ,\\
\delta v_6&=- 2R_c \varphi_0 \ ,\\
\delta h_5&=-2 R_c \psi_0 \ , \\
\delta h_6&=2 R_c \psi_0 \ .
\end{split}
\end{equation}
The equations of motion are solved in the form of plane waves,
\begin{equation}\label{eq_planewavesphere}
\mathbf{V}_\text{i}^{SH}= 
\begin{pmatrix}
\Phi_\text{i}(x,y,t) \\ \Psi_\text{i} (x,y,t)  \\ w_\text{i}(x,y,t) 
\end{pmatrix}
= 
\mathbf{v}^{SH} 
e^{\text{j} \omega t - 2\text{j} q_x  x_{\text{i}} -2\text{j} q_y  y_{\text{i}}} \ ,
\end{equation}
with the new variable $\Phi=R_c\varphi$ and $\Psi=R_c\psi$ and $\mathbf{v}^{SH} =\begin{pmatrix}  A_{\Phi} \\ A_{\Psi} \\ A_w \end{pmatrix}$ the amplitude vector.\\
Equation~(\ref{eq_planewavesphere}) is then developed around the equilibrium position ($x_0$,$y_0$) of the central particle, \\$\mathbf{V}_{\text{i}}^{SH}$=$\mathbf{v}^{SH} e^{\text{j} \omega t - 2\text{j} q_x x_0 -2\text{j} q_y y_0} e^{ -2\text{j} q_x \Delta x_{\text{i}} -2\text{j} q_y \Delta y_{\text{i}}}$, where $\Delta x_{\text{i}}=x_{\text{i}}-x_0$ and $\Delta y_{\text{i}}=y_{\text{i}}-y_0$ are the relative coordinates between the central particle and the ${\text{i}}$-th particle, and $\omega$ is the angular frequency.\\ \\
Finally, the substitution of Eq.~(\ref{eq_planewavesphere}) into the set of Eqs.~(\ref{eq_motionsphere}),~(\ref{deltasphere}) and~(\ref{deltasphere2}), leads to the eigenvalue problem,
\begin{equation} \label{eqC4:detSH}
\mathbf{S}^{SH} \mathbf{v}^{SH} = 0\ ,
\end{equation}
where $\mathbf{S}^{SH}$ is the dynamical matrix defined by
\begin{equation} \label{matrixSsphere}
\mathbf{S}^{SH}=
\left(\begin{array}{ccc}
\scriptstyle -p (\cos^2 q_y +1 +p_B  \sin^2 q_y )+\Omega^2 &\scriptstyle 0 &\scriptstyle -\text{j} p\sin q_y \cos q_y  \\\scriptstyle 0 &\scriptstyle -p (\cos^2 q_x +1 +p_B \sin^2 q_x)+\Omega^2 &\scriptstyle \text{j} p \sin q_x \cos q_x \\ \scriptstyle \text{j}  \sin q_y \cos q_y &\scriptstyle -\text{j}  \sin q_x \cos q_x &\scriptstyle -\sin^2 q_x -\sin^2 q_y +\Omega^2
\end{array}\right) \ .
\end{equation}
Nontrivial solutions of Eq.~(\ref{eqC4:detSH}) require that
\begin{equation} \label{C4eq:detSH_b}
|S_{j,i}^{SH}|=0 \ .
\end{equation}
For a given set of parameters $p$, $\eta$ and $p_B$ and wave number $q_x$, Eq.~(\ref{C4eq:detSH_b}) relates the frequency $\Omega$ and the wave number $q_y$. Eq.~(\ref{C4eq:detSH_b}) can be written in the form of a quadratic  equation for $Y=\sin^2 q_y$ or of a cubic equation for $\Omega^2$, see appendix~(\ref{appendixSH}). Note that for this particular structure, three modes exist, but only two distinct wave numbers $q_y$ with negative imaginary part correspond to a given frequency. This particularity results from the absence of $q_y$ in the second line of the dynamical matrix $\mathbf{S}^{SH}$, Eq.~(\ref{matrixSsphere}). From this second line, the following relation between the rotational amplitude $A_{\Psi}$ and the translational amplitude $A_w$ is derived
\begin{equation} \label{eq:ampSH}
A_{\Psi}=\dfrac{\text{j}p \sin q_x \cos q_x}{p(\cos^2 q_x+1+p_B\sin^2q_x-\Omega^2)}A_{w} \ .
\end{equation}
Equation~(\ref{eq:ampSH}) does not depend on $q_y$, then, the rotational amplitude $A_{\Psi}$ has the same distribution along the $y$-axis as the translational amplitude $A_{w}$. From the physics point of view, the similar distribution of modes $\Psi$ and $w$ in depth results from the fact that interaction between them, described by Eq.~(\ref{eq_motionsphere}c), includes only the beads of the same horizontal plane at a particular depth $y$ and not  the beads of different horizontal layers. At all point $x$ and $y$ in the crystal, the transversal and rotational components of the modes are assumed to be of the following form
\begin{equation}
\begin{pmatrix}
\Phi \\ \Psi \\ w
\end{pmatrix}
_{l,n}
=
\begin{pmatrix}
A_{\Phi} \\ A_{\Psi} \\ A_w
\end{pmatrix}
e^{\text{j} \omega t -2\text{j} q_x\,l -  2\text{j} q_y\,n}
=A_w
\begin{pmatrix}
 \alpha \\ \beta \\ 1 
 \end{pmatrix}
e^{\text{j} \omega t -2\text{j} q_x\,l -  2\text{j} q_y\,n} \ ,
\end{equation}
with $\alpha=\dfrac{-\text{j}p\sin q_y \cos q_y}{p( \cos^2 q_y + 1 +p_B\sin^2 q_y)-\Omega^2} $ and $\beta=\dfrac{\text{j}p\sin q_x \cos q_x}{p( \cos^2 q_x +1+p_B\sin^2 q_x)-\Omega^2}$ .\\

\begin{figure}[!ht]  
\centering
\includegraphics[scale=0.4]{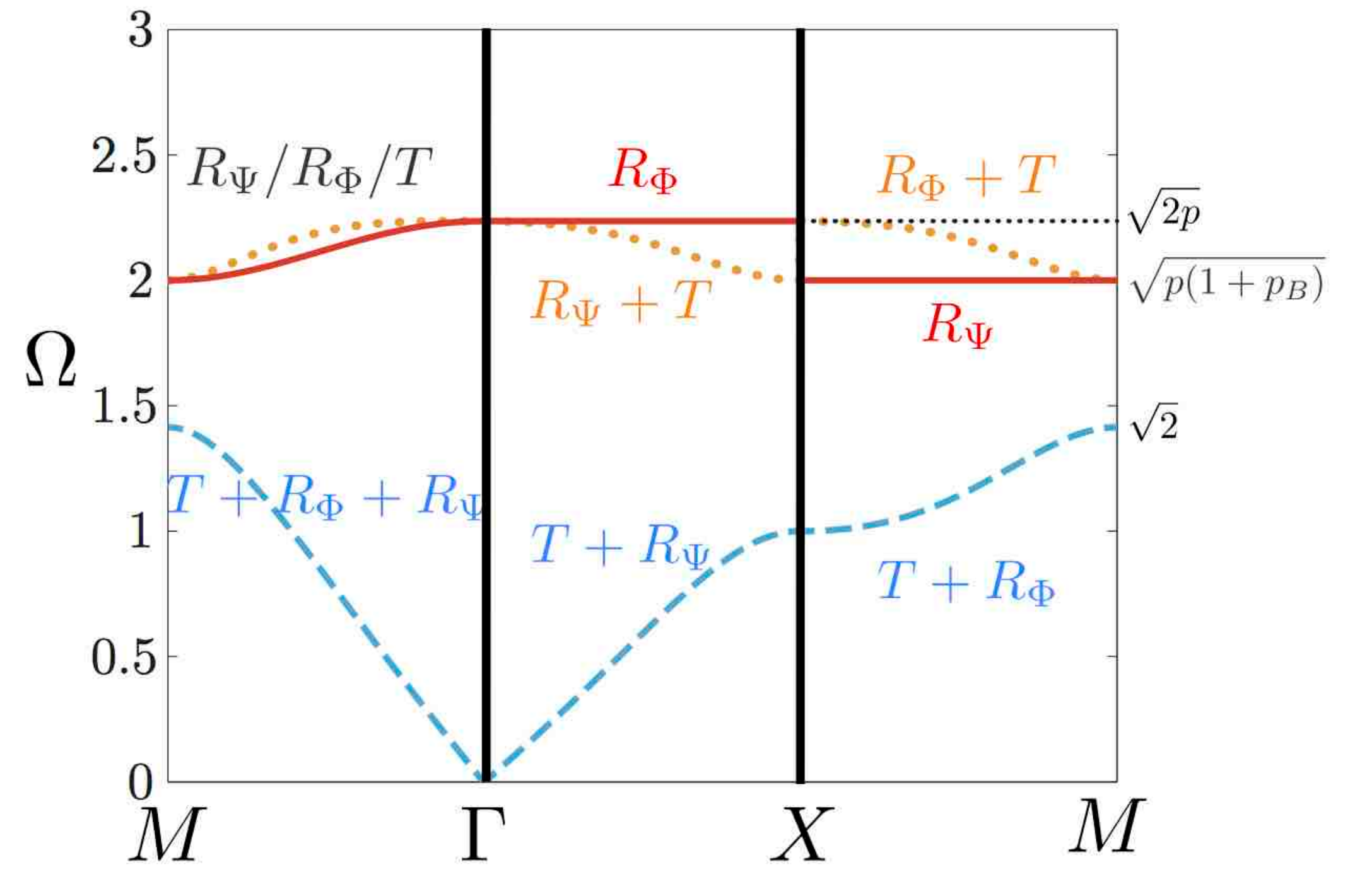}  
\caption{(Color online) Dispersion curves for a crystal made of filled spheres ($p=2.5$) and $p_B=0.6$. $T$ represents the translational motion along the $z$-axis, $R_{\Phi}$ (respectively $R_{\Psi}$) represents the rotational motion around the $x$-axis (respectively $y$-axis).}
\label{fig:dispersioncurves}
\end{figure}

Figure~\ref{fig:dispersioncurves} presents the dispersion curves along the Brillouin zone $M\Gamma XM$ for a filled sphere ($p=2.5$). Three modes propagate in the structure. The eigenmodes of the granular phononic crystal consist of three components, the translational motion $T$, the rotational motion $R_{\Phi}$ around the $x$-axis and the rotational motion $R_{\Psi}$ around the $y$-axis. The blue dashed curves correspond to coupled transverse/rotational modes with a predominance of translation, orange dotted curves correspond to coupled transverse/rotational modes with a predominance of rotation. In the case of a propagative wave in the $\Gamma X$ (respectively $XM$) direction, a mode called $R_{\Phi}$-mode (respectively $R_{\Psi}$-mode) and shown in red line appears uncoupled from the mixed modes ($T+R_{\Psi}$) (respectively ($T+R_{\Phi}$)). Along the $\Gamma M$ direction, a coupled transverse/rotational ($R_{\Phi} + R_{\Psi} + T$) mode, plotted in red line, is propagating with the same predominance for all $q_{x,y}$.

\begin{figure}[!ht]  
\centering
\includegraphics[scale=0.35]{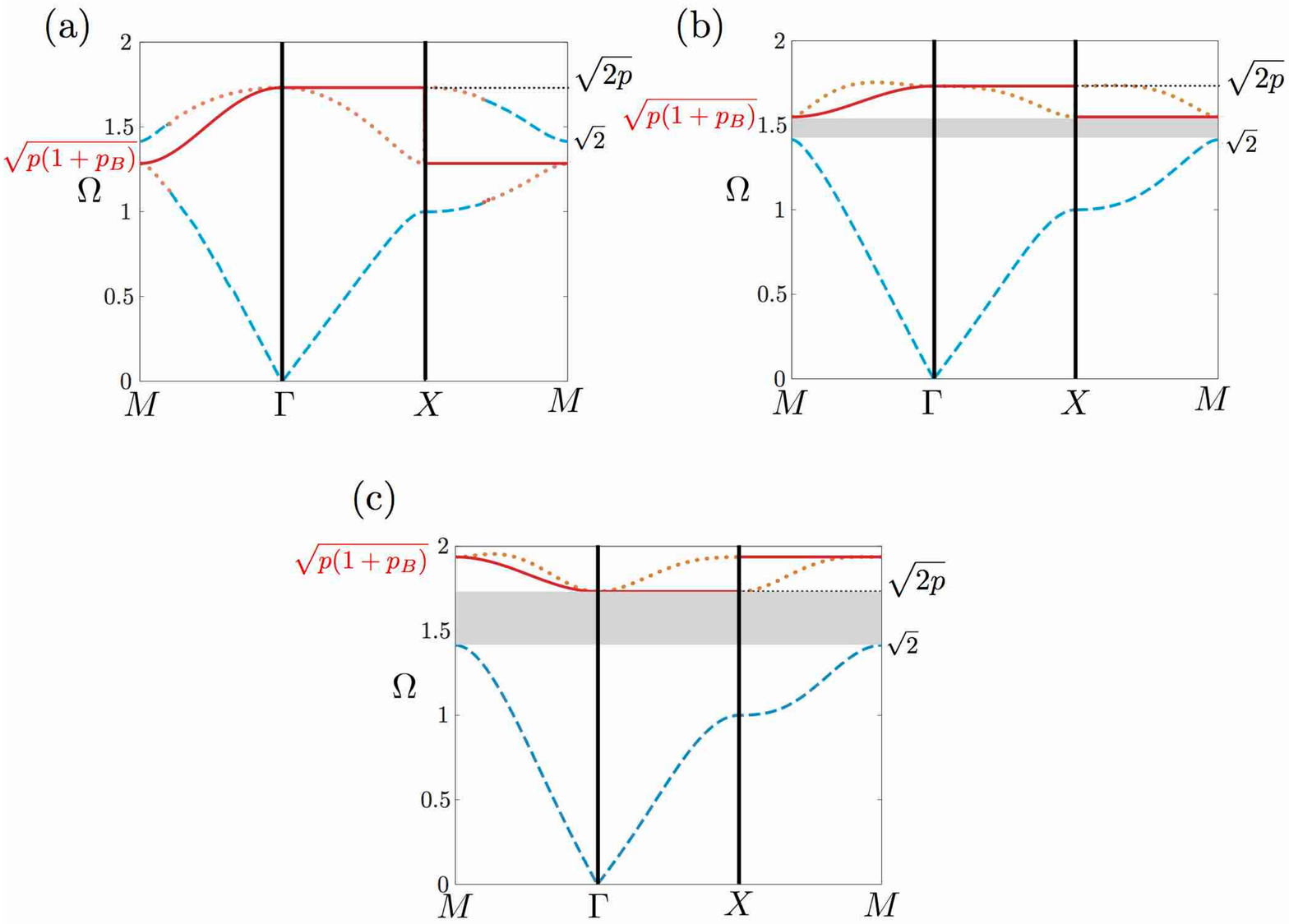}  
\caption{(Color online) Description of the dispersion curves for $p<2$ ($p=1.5$) and (a) $0\leq p_B<2/p-1$ ($p_B=0.1$), (b) $2/p-1<p_B<1$ ($p_B=0.6$), (c) $p_B>1$ ($p_B=1.5$). The complete band gap is shown by the gray shaded area.}
\label{fig:structure_sphere15}
\end{figure}

\begin{figure}[!ht]  
\centering
\includegraphics[scale=0.35]{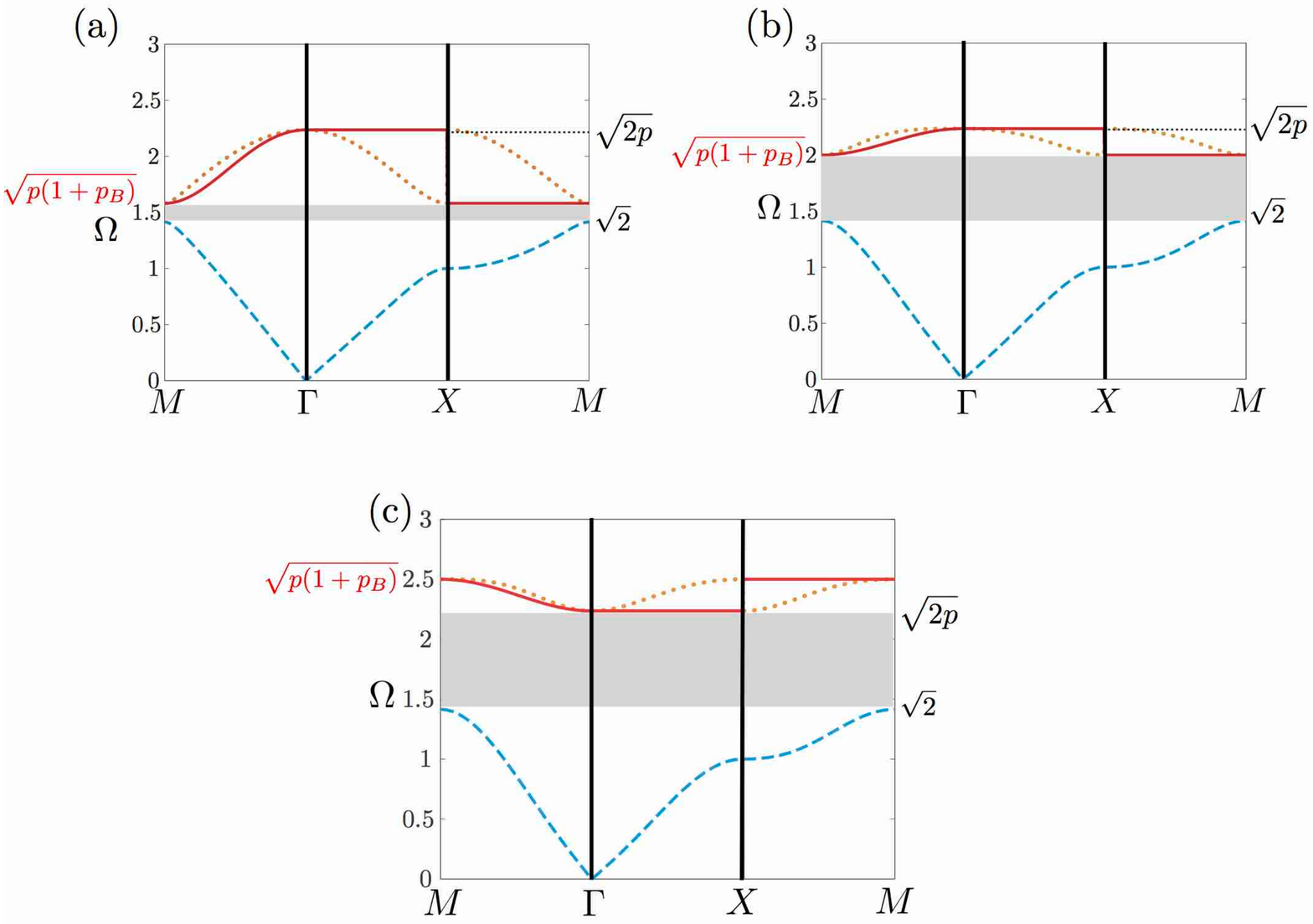}  
\caption{(Color online) Description of the dispersion curves for $p>2$ ($p=2.5$) and (a) $0\leq p_B<2/p-1$ ($p_B=0$), (b) $2/p-1<p_B<1$ ($p_B=0.6$), (c) $p_B>1$ ($p_B=1.5$). The complete band gap is shown by the  gray shaded area.}
\label{fig:structure_sphere25}
\end{figure}

Figures~\ref{fig:structure_sphere15} and~\ref{fig:structure_sphere25} present the evolution of the dispersion curves as a function of the bending rigidity parameter for empty ($p=1.5$) and filled ($p=2.5$) spheres, respectively. By increasing the bending rigidity parameter $p_B$, the modes with a predominance of rotation are shifted to high frequencies. A complete band gap, i.e., a band gap in all the directions of the Brillouin zone, exists when $p_B>2/p-1$.

\subsection{Boundary conditions for SH-type SAWs propagating at the (010) surface along [100] direction}
We study here the SH-type SAWs on the mechanically free surface of the granular crystal, which is normal to the $y$-axis, i.e., on (010) surface. As developed in section~\ref{ch4:sectBC} for Rayleigh-type SAWs, the boundary conditions are formed by removing all the particles on one side of the boundary layer. For this crystal, the mechanically free boundary conditions lead to the following equations\\
\begin{itemize}
\item No shear spring elongation,
\begin{equation}\label{eqC4:BC_GX_SH1}
 w_0-w_4 -(\Phi _0 +\Phi _4)=0 \ .
 \end{equation}

\item No rotation, able to activate bending rigidity,
\begin{equation}\label{eqC4:BC_GX_SH2}
\Phi _4 - \Phi _0=0 \ .
\end{equation}
\end{itemize}

As demonstrated in the previous section, only two wave numbers exist for one frequency, noted $q_{y_1}$ and $q_{y_2}$, because the amplitude of the rotational component $A_{\Psi}$ has the same distribution as the amplitude of the transverse component $A_w$  along the $y$-axis. The boundary conditions are then solved accounting for these two modes. If $A_{w_{i}}$, with $i=1,2$, are the amplitude of the transversal component of these two modes, the amplitudes of the displacement and rotations can be written in the form
\begin{subequations}\label{eq:ampl}
\begin{gather}
\Phi_{l,n}=\sum\limits_{i=1}^{2} A_{w_{i}} \, \alpha_i \, e^{\text{j}\omega t} e^{-2\text{j}l q_x}e^{-2\text{j}n q_{y_{i}}} \ ,\\
\Psi_{l,n}=\sum\limits_{i=1}^{2} A_{w_{i}} \, \beta \, e^{\text{j}\omega t} e^{-2\text{j}l q_x}e^{-2\text{j}n q_{y_{i}}} \ ,\\
w_{l,n}=\sum\limits_{i=1}^{2} A_{w_{i}} \, e^{\text{j}\omega t} e^{-2\text{j}l q_x}e^{-2\text{j}n q_{y_{i}}} \ . 
\end{gather}
\end{subequations}
The substitution of the amplitudes given in Eqs.~(\ref{eq:ampl}) into the boundary conditions~(\ref{eqC4:BC_GX_SH1}) and~(\ref{eqC4:BC_GX_SH2}) leads to
\begin{subequations}
\begin{gather}
\sum\limits_{i=1}^{2} A_{w_{i}} \,(1-e^{2\text{j}q_{y_i}})-A_{w_{i}} \alpha_i (1+e^{2\text{j}q_{y_i}}) =0 \ ,\\
\sum\limits_{i=1}^{2} A_{w_{i}}\, \alpha_i(e^{2\text{j}q_{y_i}}-1)=0 \ ,
\end{gather}
\end{subequations}
which can be written in the following form
\begin{equation}\label{eqC4:probS2SH}
\mathbf{S_2}^{SH} \mathbf{v_2}^{SH} =0 \ ,
\end{equation}
with $\mathbf{v_2}^{SH} =\begin{pmatrix}A_{w_{1}} \\ A_{w_{2}}  \end{pmatrix}$ and
\begin{equation} \label{matrixS2SH}
\mathbf{S_2}^{SH}=
\left(\begin{array}{cc}
 (1-e^{2\text{j}q_{y_1}})- \alpha_1 (1+e^{2\text{j}q_{y_1}}) & (1-e^{2\text{j}q_{y_2}})- \alpha_2 (1+e^{2\text{j}q_{y_2}}) \\
\alpha_1(e^{2\text{j}q_{y_1}}-1) & \alpha_2( e^{2\text{j}q_{y_2 }}-1) 
 \end{array}\right) \ ,
\end{equation}
with $\alpha_i=\dfrac{-\text{j}p\sin q_{y_i} \cos q_{y_i}}{p (\cos^2 q_{y_i} +1+p_B\sin^2 q_x)-\Omega^2} $ with $i=1,2$.\\\\
The following equation must be satisfied in order to have nontrivial solutions of Eq.~(\ref{eqC4:probS2SH})
\begin{equation} \label{eqC4:det2SH}
|S^{SH}_{2_{j,i}}|=0 \ .
\end{equation}

\subsection{Surface mode description}
\begin{figure}[!ht]
\centering
\includegraphics[scale=0.45]{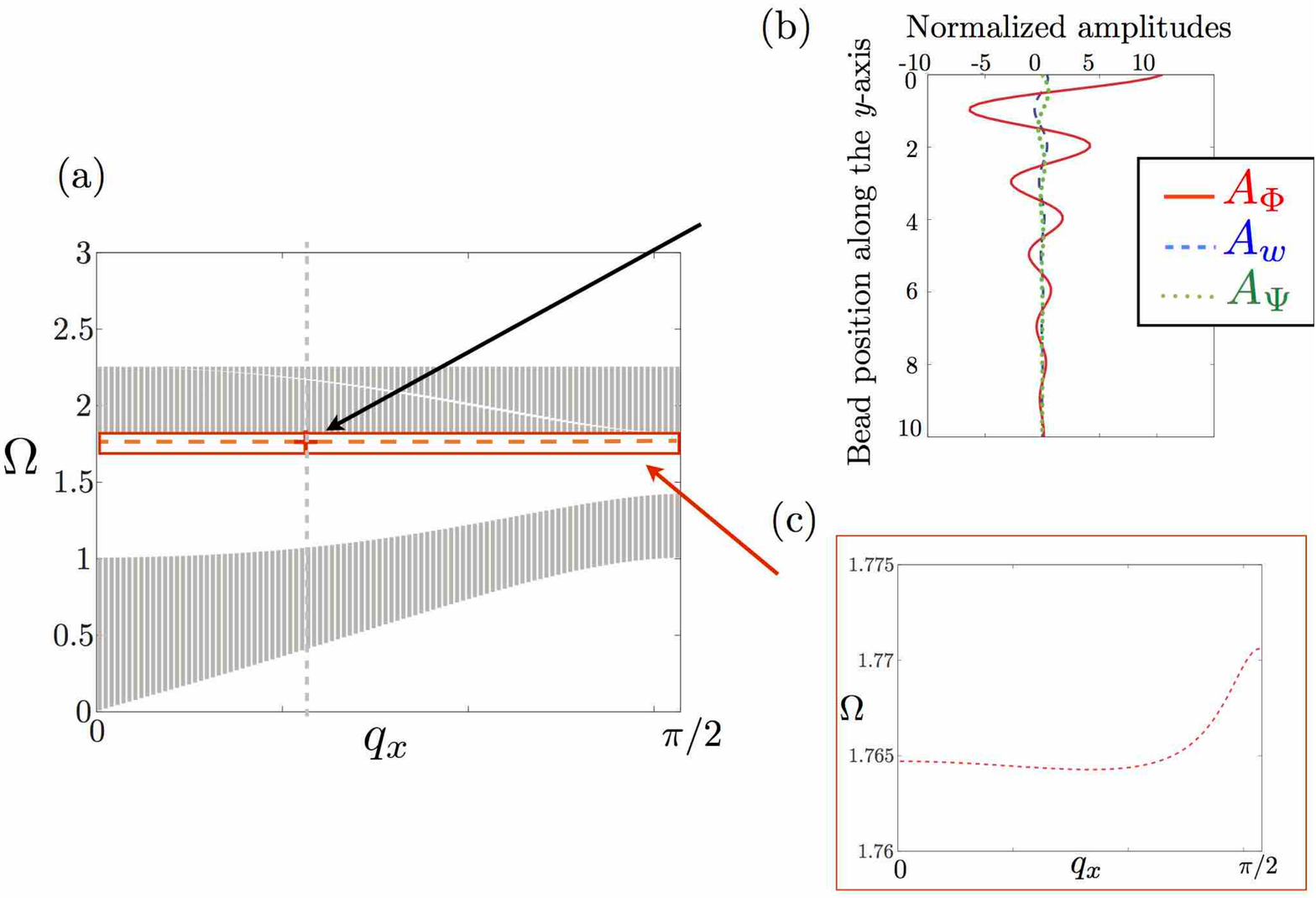}
\caption{(Color online) (a) Dispersion curves along the $q_x$ direction of the crystal for $p=2.5$ and (a) $p_B=0.3$. The shaded areas represent the projected bulk bands along [100], i.e., $x$ direction. The dashed orange curves represent the surface mode. (b) Discrete displacement profiles of the surface modes. (c) Zoom on the surface mode.}
\label{fig:dispSurfaceSphere}
\end{figure}

Surface waves are calculated for fixed sets of parameters $q_x$, $p$ and $p_B$ by simultaneous fulfilment of Eqs.~(\ref{C4eq:detSH_b}) and Eqs.~(\ref{eqC4:det2SH}). In absence of bending rigidity ($p_B=0$), no SH-type surface waves exist in the crystal, see developement in appendix~\ref{appendix:SHpB0}. When $p_B\geq0$, one surface mode exists in this granular phononic crystal. After some reduction of Eq.~(\ref{eqC4:det2SH}), the analytical form of the surface mode frequency is found
\begin{equation}\label{eq:ModeS_SH}
\Omega_S^2= p+\dfrac{p_Bp}{2}-\dfrac{1}{2}p_B p \dfrac{\cos(q_{y_1}-q_{y_2})}{\cos(q_{y_1}+q_{y_2})} \ ,
\end{equation}
with $q_{y_1}$ and $q_{y_2}$ solutions of Eq.~(\ref{eqC4:SySH}) (annexe~\ref{appendixSH}). This mode is plotted in orange dashed line in Fig.~\ref{fig:dispSurfaceSphere}(a) in the case of filled sphere ($p=2.5$) and with a bending rigidity parameter $p_B=0.3$. This surface mode is slightly dispersive, Fig.~\ref{fig:dispSurfaceSphere}(c). For various values of the parameters, it is localized around one frequency. The possible reason for this strong localization in the frequency domain is the weak coupling of rotational motion $R_\Phi$ with the other motions. 
According to Eqs.~(\ref{eq:ampl}), the amplitudes of the discrete displacement and rotations of the transversal $w_{l,n}$, rotational $\Phi_{l,n}$ and $\Psi_{l,n}$ components of the surface modes as a function of the particle position $(l,n)$ in the crystal can be determined by combining the two evanescent modes
\begin{equation}
\begin{pmatrix}
w_{l,n} \\ \Phi_{l,n}\\ \Psi_{l,n}
 \end{pmatrix}
 =
 A_{w_{1}}\left[  \begin{pmatrix} 1 \\ \alpha_1 \\ \beta \end{pmatrix} e^{\text{j}\omega t} e^{-2\text{j} l q_x}e^{-2\text{j}n q_{y_{1}}} + Z \begin{pmatrix} 1 \\ \alpha_2 \\ \beta \end{pmatrix} e^{\text{j}\omega t} e^{-2\text{j}l q_x}e^{-2\text{j}n q_{y_{2}}} \right] \ ,
\end{equation}
with $Z=\dfrac{A_{w_{2}}}{A_{w_{1}}}=-\dfrac{\alpha_1(1-e^{-2\text{j} q_{y_1}})}{\alpha_2(1-e^{-2\text{j} q_{y_2}})}$.
As illustrated in Fig.~\ref{fig:dispSurfaceSphere}(b), due to the symmetry and configuration of the crystal, this particular mode has mainly a rotational $A_{\Phi}$ component. The decays of the amplitudes are a combination of a monotonously decaying function and a decaying function with few oscillations.

\subsection{SH-type SAWs propagating at the (110) surface along [1$\bar{1}$0] direction}

\begin{figure}[!ht]
\centering
\includegraphics[scale=0.45]{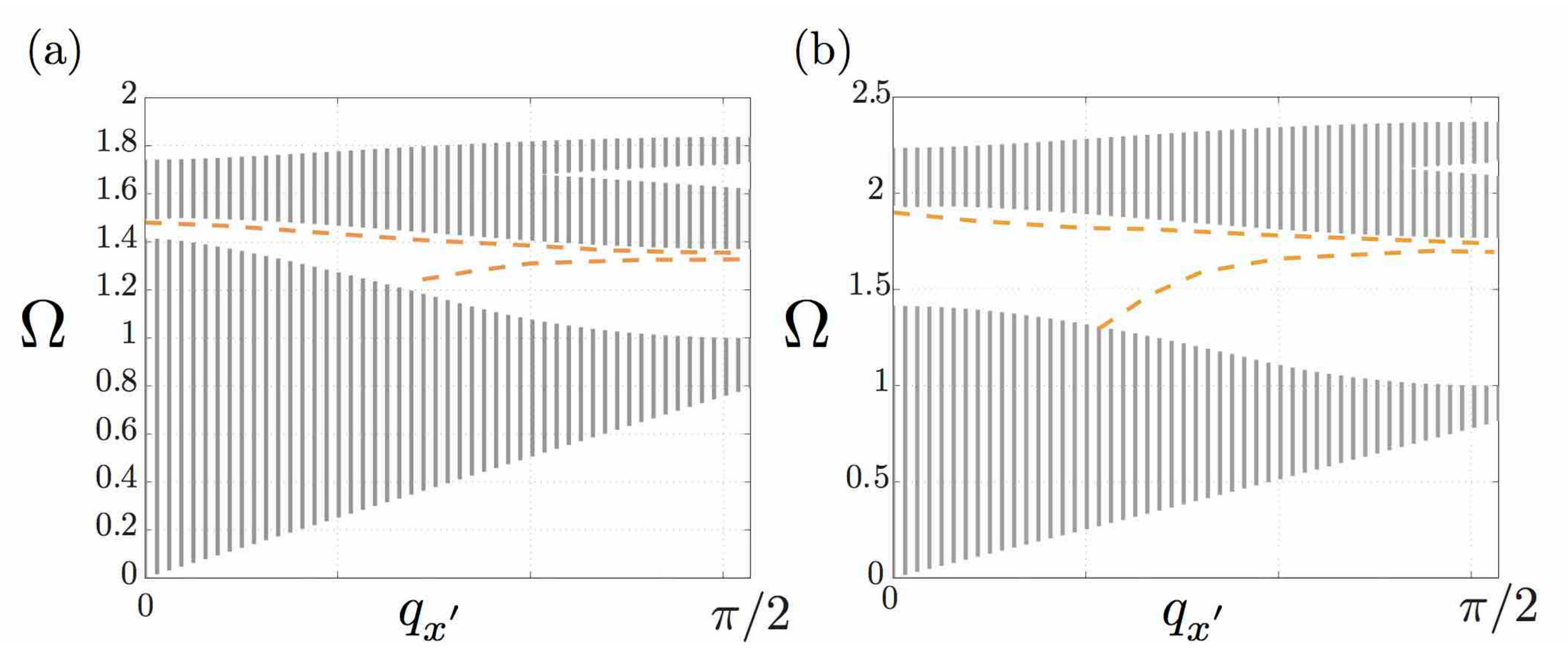}
\caption{(Color online) Dispersion curves along the diagonal direction of the crystal for $p_B=0.5$ and (a) empty sphere ($p=1.5$) and (b) filled sphere ($p=2.5$). The shaded areas represent the projected bulk bands along [1$\bar{1}$0] and the dashed orange curves represent the surface modes.}
\label{fig:SAW_diag_SH}
\end{figure}

SH-type SAWs propagating at the (110) surface along [1$\bar{1}$0] direction are investigated in this part. The  surface position is the same as the one presented in Fig.~\ref{figC4:diag}. The eigenvalue problem resulting from the substitution of the plane wave solutions into the equation of motions is
\begin{equation} \label{eqC4:detS_diagSH}
\mathbf{S_{\textrm{diag}}^{SH}} \mathbf{v^{SH}} = 0 \ ,
\end{equation}
with $\mathbf{v^{SH}} =\begin{pmatrix} A_{\Phi} \\ A_{\Psi} \\ A_{w} \end{pmatrix}$ and
\begin{equation} \label{eqC4:matrixS_diagSH}
\mathbf{S_{\textrm{diag}}^{SH}}=
\left(\begin{array}{ccc}
\scriptstyle -\frac{p}{2}\cos q_x^{'} \cos q_y^{'}(1-p_B)-\frac{p_B p}{2}-\frac{3p}{2}+\Omega^2 & \scriptstyle -\frac{ \textrm{j}}{2}p \sin q_x^{'} \sin q_y^{'} &\scriptstyle -\frac{1}{\sqrt{2}} p\cos q_x^{'} \sin q_y^{'} \\
\scriptstyle -\frac{1}{2} p \sin q_x^{'} \sin q_y^{'} & \scriptstyle -\frac{p}{2}\cos q_x^{'} \cos q_y^{'}(1-p_B)-\frac{p_B p}{2}-\frac{3p}{2}+\Omega^2  & \scriptstyle\frac{\textrm{j}}{\sqrt{2}}p \sin q_x^{'} \cos q_y^{'} \\
\scriptstyle \frac{\textrm{j} }{\sqrt{2}}  \cos q_x^{'} \sin q_y^{'} & \scriptstyle-\frac{\textrm{j} }{\sqrt{2}}\sin q_x^{'} \cos q_y^{'} & \scriptstyle -1+\cos q_x^{'} \cos q_y^{'}+\Omega^2
\end{array}\right) .
\end{equation}
Mechanically free boundary conditions are applied at the surface, i.e., the total forces of beads $1$ and $4$ acting on bead $0$ vanish. The amplitudes $A_{\Phi_i}$, $A_{\Psi_i}$ and $A_{w_i}$ corresponding to a particular $q^{'}_{y_i}$ can be determined by
\begin{equation}
\dfrac{A_{\Phi_i}}{\chi_i}=\dfrac{A_{\Psi_i}}{\epsilon_i}=\dfrac{A_{w_i}}{\zeta_i}=\Lambda_i,
\end{equation}
where $\chi_i$, $\epsilon_i$ and $\zeta_i$ are the cofactors of all row of the dynamical matrix~(\ref{eqC4:matrixS_diagSH}) associated with $q^{'}_{y_i}$ ($i=1,2,3$) and where the $\Lambda_i$ are to be determined from the boundary conditions. \\
Hence, the general solution is
\begin{equation} \label{eq:exp_cof}
\begin{pmatrix}
\Phi \\  \Psi \\ w  
\end{pmatrix}
=
\sum\limits_{i=1}^{3} (\chi_i, \epsilon_i,  \zeta_i) \, \Lambda_i \, e^{\text{j} \omega t - \text{j} q^{'}_x \, x^{'} -\text{j}  q^{'}_y\,  y^{'}}.
\end{equation}
Substituting Eq.~(\ref{eq:exp_cof}) into the boundary condition system leads to
\begin{equation}\label{eq:EVP2SH}
\sum\limits_{i=1}^{3} S_{2\textrm{diag}_{j,i}}^{SH} \, \Lambda_i =0 \quad (i,j=1,2,3),
\end{equation}
where
\begin{equation}
\begin{split}
S^{SH}_{2\textrm{diag}_{1,i}}&=\chi_i (1-\cos q_x^{'} e^{\textrm{j}q_{y_i}^{'}}) - \epsilon_i \dfrac{1}{\sqrt{2}}(1+\cos q_x^{'} e^{\textrm{j}q_{y_i}^{'}}) + \dfrac{\textrm{j}}{\sqrt{2}}\zeta_i \sin q_x^{'} e^{\textrm{j}q_{y_i}^{'}},\\
S^{SH}_{2\textrm{diag}_{2,i}}&=\chi_i (1-\cos q_x^{'} e^{\textrm{j}q_{y_i}^{'}}) - \epsilon_i \left(\dfrac{1}{\sqrt{2}}(1+\cos q_x^{'} e^{\textrm{j}q_{y_i}^{'}})+\sqrt{2} p_B(1-\cos q_x^{'} e^{\textrm{j}q_{y_i}^{'}})\right)+ \dfrac{\textrm{j}}{\sqrt{2}} \zeta_i \sin q_x^{'} e^{\textrm{j}q_{y_i}^{'}} ,\\
S^{SH}_{2\textrm{diag}_{3,i}}&=\textrm{j}\chi_i \sin q_x^{'} e^{\textrm{j}q_{y_i}^{'}} +\dfrac{\textrm{j}}{\sqrt{2}} \epsilon_i \sin q_x^{'} e^{\textrm{j}q_{y_i}^{'}}- \zeta_i \left(\dfrac{1}{\sqrt{2}}(1+\cos q_x^{'} e^{\textrm{j}q_{y_i}^{'}})+\sqrt{2}p_B(1-\cos q_x^{'} e^{\textrm{j}q_{y_i}^{'}})\right).
\end{split}
\end{equation}
Surface waves can then be obtained by simultaneous fulfilment of Eqs~(\ref{eqC4:detS_diagSH}) and~(\ref{eq:EVP2SH}).\\\\
The required condition for the existence of surface waves is the presence of bending rigidity ($p_B>0$). Figures~\ref{fig:SAW_diag_SH}(a) and (b) illustrate the obtained surface modes in the case of $p_B=0.5$ and empty sphere ($p=1.5$) and filled sphere ($p=2.5$), respectively. Two branches are found in the gap between the two first propagation bands.\\

Another important feature of the predicted surface acoustic wave, which is described by Eq.~(\ref{eq:ModeS_SH}), is its existence only in the presence of bending rigidity of the contacts. If the direct bending-type interactions between the rotations of the beads are neglected this surface mode transforms into the purely vibrational mode with frequency equal to $\sqrt{p}$ and zero group velocity. This is a non-negligible bending rigidity of the contact which induces propagation of this surface acoustic wave. It is worth noticing here that earlier~\cite{Pichard12,Pichard14} it was demonstrated that some zero-energy, i.e., $w=0$, modes of granular phononic crystals become propagative due the the bending-type interactions between the beads.

\section{Comparison with the Cosserat theory} \label{ch4:sectCosserat}
\subsection{Brief introduction to the Cosserat theory}
One hundred years ago, the Cosserat brothers developed a continuum elasticity theory accounting for the rotational \textit{dofs} of point bodies (infinitesimal particles) constituting deformable solids~\cite{Cosserat09}. Currently, this theory is known as the Cosserat theory, and the related and advanced theories are known as theories of Cosserat continuum or as theories of micropolar continuum~\cite{Eringen99}. In the Cosserat theory, each material point has six \textit{dofs}, three of which correspond to the translations as in the classical theory of elasticity, and the three others correspond to rotations. The stress tensor is asymmetric and an additional couple-stress tensor is introduced, which plays the analogous role for torques than the stress tensor plays for forces. The theory predicts a contribution of rotations to the dispersion of the shear elastic wave velocity as well as the existence of additional rotational bulk modes. More information on the Cosserat theory can be found in the Supplemental Material~\cite{PichardSM}.\\

The additional effects predicted by this theory have never been observed experimentally and have been subjected to criticism~\cite{Eringen99,Pasternak00}. More recently, the rotational modes have been revealed experimentally in a 3D granular phononic crystal~\cite{Merkel11}. A theoretical comparison of the bulk waves in homogenized three-dimensional granular phononic crystals with those in the Cosserat continuum has demonstrated that the Cosserat theory does not account for all the influences of the material inhomogeneity on its elastic behavior. To go further in these previous conclusions, a theoretical comparison of surface waves in the granular crystals with those in the reduced Cosserat theory is performed below.

\subsection{Comparison of SAWs in granular crystals and in reduced Cosserat medium}\label{sec:C4Coss}
In continuum elasticity, the Cosserat theory~\cite{Cosserat09} and its various extensions~\cite{Mindlin64,Toupin64,Eringen68,Eringen99,Askar86} introduce the rotational \textit{dofs} of an elementary volume for modeling wave phenomena in micro-inhomogeneous materials. In these theories the motion of an elementary "point" is characterized, in addition to the mechanical displacement vector $\vec{u}$, by the vector of  mechanical rotation (angle) $\vec{\theta}$. In the simplest case of the so-called reduced Cosserat continuum, in addition to Lam\'e moduli $\lambda$ and $\mu$, just a single modulus $\alpha$, coupling the displacements and rotations, is introduced. For the harmonic waves of cyclic frequency $\omega$ with wave vector $\vec{k}$ the coupled equations of the reduced Cosserat continuum are~\cite{Kulesh09} 
\begin{equation}\label{eq:Coss1}
\rho\, \omega^2 \vec{u}=(\lambda+2 \mu) \vec{k} (\vec{k} \vec{u}) -(\mu+\alpha) \left[\vec{k}[\vec{k} \vec{u}]\right] + \text{j}2\alpha [\vec{k} \vec{\theta}],
\end{equation}

\begin{equation}\label{eq:Coss2}
J \omega^2 \vec{\theta}=4 \alpha \vec{\theta} + 2\text{j}\alpha [\vec{k}\vec{u}],
\end{equation}
where $\rho$ is the density and $J$ denotes the density of the moment of inertia.\\\\
From Eq.~(\ref{eq:Coss2}) it follows that the modulus $\alpha$ together with $J$ control the rotational resonance frequency $\omega_0$ of elementary volumes
\begin{equation}\label{eq:Coss3}
\vec{\theta}=\text{j}\dfrac{\omega_0^2}{2(\omega^2-\omega_0^2)}[\vec{k}\vec{u}] \quad \textrm{with} \quad \omega_0=2\sqrt{\dfrac{\alpha}{J}}.
\end{equation}
Subsitution of Eq.~(\ref{eq:Coss3}) into Eq.~(\ref{eq:Coss1}) gives
\begin{equation}\label{eq:Coss4}
\rho\, \omega^2 \vec{u}=(\lambda+2 \mu) \vec{k} (\vec{k}\vec{u})-\left( \mu+\alpha \dfrac{\omega^2}{\omega^2-\omega_0^2}\right) \left[\vec{k}[\vec{k} \vec{u}] \right] .
\end{equation}

Eq.~(\ref{eq:Coss4}) demonstrates that the acoustic waves with the longitudinal polarization of displacement, $[\vec{k}\vec{u}]=0$, are not modified in comparison with classical elasticity theory, while the modification of the waves with transverse polarization of the displacement, $(\vec{k}\vec{u})=0$, takes place as in a metamaterial with resonant inclusions~\cite{Liu00,Liu11X}. Moreover, because local resonances of the reduced Cosserat metamaterial are rotational, they modify the effective modulus of the "metamaterial" and not its density (see~\cite{Liu11X,Gusev14} and references therein). In accordance with Eq.~(\ref{eq:Coss4}), the dispersion relation of the coupled transverse/rotational modes, $(\vec{k}\vec{u})=0$, can be presented in the form 
\begin{equation}
k^2=\dfrac{\omega^2}{C_T^2}\dfrac{1}{1+\dfrac{\alpha}{\mu}\dfrac{\omega^2}{\omega^2-\omega_0^2}}.
\end{equation}
It describes two branches, the $RT$ branch at frequencies above $\omega_0$ and $TR$ branch at frequencies below $\omega_1=\dfrac{\mu}{\mu+\alpha}\omega_0=\left(\dfrac{C_{TR}}{C_{RT}}\right)^2\omega_0$, separated by the band gap, $\omega_1 \leq \omega \leq \omega_0$, where there is no propagative mode because of the negative effective modulus of reduced Cosserat medium. In the above formula $C_{TR}$ denotes the velocity of the lower mode when $\omega \rightarrow 0$ ($k \rightarrow 0$), while $C_{RT}$ denotes the velocity of the upper mode when the mode extends at $\omega \rightarrow \infty$ $(k \rightarrow \infty$).\\
\\ The SAWs in reduced Cosserat continuum have been studied in details quite recently~\cite{Kulesh09}. The theory predicted that coupling of the $TR$ and $RT$ bulk modes with longitudinal bulk acoustic waves at the mechanically free surface leads to the existence of two branches of Rayleigh type SAWs. The lower branch of SAWs is below the $TR$ bulk branch. The upper branch of Rayleigh-type SAWs is below the dispersion curves of $RT$ and $L$ bulk branches and above the low edge, $\omega=\omega_1$, of the bulk band gap. Note that surface waves can be propagative at frequencies forbidden for bulk $TR$ and $RT$ wave propagation. Because of the above formulated requirements, the minimum possible wave number, $k$, for the upper branch of Rayleigh-type SAW cannot be smaller than $k_{min}=\omega_1/C_L$, where $C_L=\sqrt{\dfrac{\lambda+2\mu}{\rho}}$ is the velocity of dispersionless bulk longitudinal waves. The theory of SAWs in granular phononic crystals, which was presented above, confirms the predictions of the reduced Cosserat theory on the possible existence of multiple Rayleigh-type surface acoustic modes (see Figs.~\ref{fig:evolutionpB} and~\ref{figC4:diag_p15_eta2}). In comparison with the Cosserat theory some of the upper branches of Rayleigh-type SAW can start at finite $\omega$ from $k=0$ (see Fig.~\ref{fig:evolutionpB}), which is forbidden in the Cosserat theory. This is related to the opening, in phononic crystal, of band gaps for the propagation of bulk longitudinal modes, which can be located near $k=0$ for finite $\omega$. Thus the limiting condition $k_{min}=\omega_1/C_L$ is lifted. However, the most important difference with the reduced Cosserat theory, from the physics point of view, is related to physical origins of wave dispersion. As it was pointed out earlier in the comparison of bulk waves in granular phononic crystal and in the Cosserat media~\cite{Merkel11}, in the former the dispersion comes both from the repulsion/hybridization of transverse and rotational motions and from multiple scattering of the waves (induced by natural spatial inhomogeneity/periodicity of phononic crystals), while in the latter the wave dispersion is caused by hybridization phenomena only. The situation with surface acoustic waves is similar. For example, in the dispersion of the lower branch of Rayleigh-type SAW at $\omega \to 0$ there are contributions due to both the interaction between different modes and also to the explicit existence in phononic crystals of a characteristic scale of spatial inhomogeneity, which is absent in the reduced Cosserat medium. 
 Thus our comparison indicates that for the correct modeling of the wave phenomena in micro-inhomogeneous media the Cosserat theories should be at least combined with higher order gradient theories of elasticity, which explicitly contain the characteristic spatial length of micro-inhomogeneity and, thus, account for wave scattering by spatial inhomogeneities.\\\\
Our theoretical analysis of the SH-type SAWs in granular crystals can be also compared with the theoretical predictions of the reduced Cosserat continuum~\cite{Kulesh09}. In fact the theory~\cite{Kulesh09} has predicted the absence of energy transporting SH-type SAWs in the reduced Cosserat continuum. It has just predicted the existence of $k$-independent vibrations, $\omega(k)=\omega_{1}=$constant, at frequency $\omega_{1}$ where the effective modulus of the reduced Cosserat medium is equal to zero. These zero-group-velocity vibrational modes can be arbitrary distributed in depth~\cite{Kulesh09}. The theory developed above for the granular phononic crystal demonstrates that these types of vibrational modes also exist and, moreover, they transform into true localized and energy carrying SAWs when bending rigidity of the contacts between the beads is taken into account (see Figs.~\ref{fig:dispSurfaceSphere} and~\ref{fig:SAW_diag_SH}). Thus our theory highlights the crucial role of bending-type interactions for some surface acoustic wave phenomena. Earlier it was demonstrated~\cite{Pichard12,Pichard14} that these types of interaction between beads induce true propagative modes in granular phononic chains~\cite{Pichard14} and in bulk granular phononic crystals~\cite{Pichard12}, which are otherwise zero-energy, i.e., $\omega=0$, modes. The theory developed above confirms that neglecting in the reduced Cosserat theory, relative to the general Cosserat theory, the direct interaction between the rotations of elementary volumes, which mathematically manifests itself in the absence of any terms containing spatial derivatives of $\vec{\theta}$ in Eq.~(\ref{eq:Coss2}), can have profound consequences in terms of wave phenomena predicted by this theory. \\\\
For the case of the SH-type surface acoustic waves the predictions of the reduced Cosserat theory of localized vibrations only, at a particular single frequency~\cite{Kulesh09}, is drastically different from the prediction by the general Cosserat theory of a true surface acoustic wave, existing in the complete domain of frequencies and wave numbers~\cite{Kulesh07_b}. The SH-type SAW dispersion curve predicted by the general Cosserat theory starts at $\omega=0$ ($k=0$) and up to $\omega=\infty$ ($k=\infty$) and is located below all the dispersion curves of bulk modes. In classical terminology it is an acoustical-type surface phonon mode. Our theory predicts possible existence of optical-type SH surface acoustic mode, which dispersion curve starts at $k=0$ from $\omega \neq 0$ (Fig.~\ref{fig:dispSurfaceSphere}). This slow mode is much closer to the surface vibrational mode predicted by the reduced Cosserat model than to the acoustical-type SH-type SAW predicted by the general Cosserat theory. Thus our theory indicates that the reduced Cosserat model can be a fruitful tool for the prediction of wave phenomena in some particular granular crystals and micro-inhomogeneous materials. The reduced Cosserat theory clearly reveals some wave phenomena which existence is so deeply hidden in the general Cosserat theory that it could be easily missed.

\section{Conclusion}

In summary, the propagation of surface waves has been analyzed at the mechanically free surface of two different granular phononic crystals. A first case with the surface along the (010) plane of a cubic crystal and surface waves propagating in the [100] direction is investigated. In the first studied granular phononic crystal, where the particles possess two translational and one rotational \textit{dofs}, generalized type Rayleigh surface waves and one pure longitudinal mode skimming along the surface direction, i.e., in [100] direction, are found. The analysis shows a nonmonotonous behavior with a zero-group velocity point for the lower frequency Rayleigh surface mode, which can be interesting for non-destructive characterization applications. The surface waves amplitude decay in depth is a combination of an exponentially decaying function with few oscillations. In the second granular phononic crystal studied, with particles possessing two rotational and one translational \textit{dofs}, one shear-horizontal type surface wave is found. This mode has the particularity to be localized around one particular frequency. We have demonstrated that the existence of this mode is due to the bending-type interaction between the rotating grains in contact and that the dispersion of this mode can be tuned by modifying the ratio of the bending and shear rigidities acting between the particles. In addition, the existence of Rayleigh-type and SH-type SAWs propagating at the (110) surface along [1$\bar{1}$0] direction is theoretically revealed. Our findings are of interest in non-destructive testing of materials and in the design of devices devoted to frequency filtering or waveguiding.  \\

A comparison of the results obtained in the granular phononic crystals with the predictions of surface waves in the Cosserat continuum is made. Our theoretical results indicate the usefullness of some simplified Cosserat theories, such as the reduced Cosserat theory, in revealing some surface wave phenomena which could be easily hidden in the frame of the general Cosserat theory. When the wavelength is shortened and becomes comparable to the particle size, the Cosserat theories are unable to describe the dispersive phononic properties. However these theories also fail to correctly account for all consequences of the material inhomogeneity in the long-wavelength limit. The obtained results confirm that the Cosserat theories do not account correctly for all effects of the material inhomogeneity on its elastic behavior, and should be combined with higher order gradient theories. The generalized elasticity theory should explicitly incorporate the spatial scale of the inhomogeneity in order to account for the multiple scattering of the waves. In perspectives, an experimental validation of the surface modes in the granular phononic crystals would be interesting; in particular in order to evaluate the physical constants appearing in the Cosserat theory. These results would allow to compare analytically discrete lattice theory and Cosserat theory and thus to identify more precisely the limits of the Cosserat theory.

\appendix
\section{Dispersion relation in the granular phononic crystal with particles possessing two translational and one rotational \textit{dofs}}\label{app:detS}
 The determinant of the eigenvalue problem~(\ref{eqC4:detS}) leads to
{\small\begin{align}\label{det1}
 & \left[-\eta \sin^2 q_x - \sin^2 q_y + \Omega^2\right] \nonumber \\
 &\times \left[\left(-\eta \sin^2 q_y -\sin^2 q_x +\Omega^2\right)\left(-p(1-\sin^2 q_x+1-\sin^2 q_y)-4p_B p (\sin^2 q_x+\sin^2 q_y)+\Omega^2\right)-p\sin^2 q_x(1-\sin^2 q_x)\right] \nonumber \\
 & -p\sin^2 q_y (1-\sin^2 q_y)\left(-\eta \sin^2 q_y-\sin^2 q_x+\Omega^2\right)=0 \ ,
\end{align}}
which can be written as a cubic equation for $Y=\sin^2 q_y$ 
\begin{align} \label{eqC4:cubicSy}
& Y^3 \left[ -4p_Bp\eta \right] \nonumber \\
& +Y^2 \left[ p\left( \eta(-1+\sin^2 q_x  +\sin^2 q_x\eta) - 4 p_B \sin^2 q_x (1+\eta+\eta^2) \right)+\Omega^2 \left( \eta-p\eta+4 p_B p(1+\eta) \right)   \right] \nonumber \\
& +Y \left[\Omega^4 \left(-1+p-4p_Bp-\eta \right)+ \Omega^2 \left(\sin^2 q_x (1+\eta^2)+p(1+2\eta+2(-1+4p_B)\sin^2 q_x(1+\eta))\right)   \right]\nonumber \\
& +Y \left[ p \sin^2 q_x \left(\eta(\sin^2 q_x-2\eta+\eta\sin^2 q_x)-4p_B \sin^2 q_x(1+\eta+\eta^2)\right)  \right] \nonumber \\
& -p \sin^4 q_x (1+4p_B \sin^2 q_x)\eta+\sin^2 q_x \Omega^2\left[\sin^2 q_x \eta+p(1+2\eta-\sin^2 q_x\eta+4p_B\sin^2 q_x(1+\eta))\right] \nonumber \\
& + \Omega^4 \left[ -\sin q_x (1+\eta) +p(-2+\sin^2 q_x-4p_B\sin^2 q_x))\right]+\Omega^6 =0 \ ,
\end{align}
and as a cubic equation for $\Omega^2$
\begin{align} \label{eqC4:cubicOm}
& (\Omega^2)^3 \nonumber \\
&+(\Omega^2)^2 \left[p(-2+\sin^2 q_x(1-4p_B)+\sin^2 q_y(1-4p_B))-(\sin^2 q_x+\sin^2q_y)(1+\eta)\right] \nonumber \\
& +\Omega^2 \,( \sin^4 q_x \eta+\sin^4 q_y\eta+\sin^2 q_x \sin^2 q_y(1+\eta^2)+p(\sin^4q_x(-\eta+4p_B(1+\eta))\nonumber\\
&+\sin^2q_y(1+2\eta-\eta \sin^2q_y+4p_B\sin^2q_y(1+\eta))+\sin^2q_x(1+2\eta+2(-1+4p_B)\sin^2q_y(1+\eta))))\nonumber \\
&-p(\eta(\sin^4q_y-\sin^4 q_x(-1+\sin^2q_y+\sin^2q_y \eta)-\sin^2q_x \sin^2q_y(\sin^2q_y-2\eta+\sin^2q_y\eta))\nonumber \\
&+4p_B(\sin^2q_x+\sin^2q_y)(\sin^4 q_x \eta+\sin^4q_y\eta+\sin^2q_x\sin^2q_y(1+\eta^2)))=0 \ .
\end{align}
Since it is a cubic equation in $\sin^2 q_y$ and in $\Omega^2$, for a given frequency $\Omega$, there are six corresponding wave numbers $q_y$. The analysis is restricted to displacement and rotational components of the surface waves whose amplitudes decrease as $n$ increases. Therefore, the attenuation of surface waves is provided by complex wave numbers with a negative imaginary part, i.e., by three of the six wave numbers given by Eq.~(\ref{eqC4:cubicSy}).
\section{Pure longitudinal mode}\label{App:planemode}
From the development of the boundary condition, the determinant Eq.~(\ref{eqC4:det2}) exhibit a pure longitudinal mode $\Omega^2=\eta \sin^2 q_x$. As developed below the term $\Omega^2-\eta \sin^2 q_x$ can be factorized in the determinant.
\begin{equation} \label{eqC4:pureL}
 \begin{split}
&|S_{2_{j,i}}|=0  \Leftrightarrow 
\left|\begin{array}{ccc}
-\beta_1 & -\beta_2 & -\beta_3 \\
 \alpha_1+\dfrac{\cos q_{y_1}}{\text{j}\sin q_{y_1}} & \alpha_2+\dfrac{\cos q_{y_2}}{\text{j}\sin q_{y_2}} &\alpha_3+\dfrac{\cos q_{y_3}}{\text{j}\sin q_{y_3}}=0 \\
1 & 1 & 1
 \end{array}\right|=0 \\
 & \Leftrightarrow 
 \left| \begin{array}{ccc}
\dfrac{1}{\eta \sin^2 q_{y_1}+\sin^2 q_x - \Omega^2}  &\dfrac{1}{\eta \sin^2 q_{y_2}+\sin^2 q_x - \Omega^2} & \dfrac{1}{\eta \sin^2 q_{y_3}+\sin^2 q_x - \Omega^2} \\
 \alpha_1+\dfrac{\cos q_{y_1}}{\text{j}\sin q_{y_1}} & \alpha_2+\dfrac{\cos q_{y_2}}{\text{j}\sin q_{y_2}} &\alpha_3+\dfrac{\cos q_{y_3}}{\text{j}\sin q_{y_3}} \\
1 & 1 & 1
 \end{array}\right|
 =0 \\
 &  \Leftrightarrow 
  \text{j}(\Omega^2-\eta \sin^2 q_x) \left| \begin{array}{ccc}
   \dfrac{\scriptstyle 1}{\scriptstyle \eta \sin^2 q_{y_1}+\sin^2 q_x - \Omega^2}  & \dfrac{\scriptstyle 1}{ \scriptstyle\eta \sin^2 q_{y_2}+\sin^2 q_x - \Omega^2} &\dfrac{\scriptstyle 1}{ \scriptstyle \eta \sin^2 q_{y_3}+\sin^2 q_x - \Omega^2} \\
  \dfrac{\scriptstyle \cos q_{y_1}}{\scriptstyle \sin q_{y_1}(\eta \sin^2 q_x + \sin^2 q_{y_1}-\Omega^2)} &   \dfrac{\scriptstyle \cos q_{y_2}}{\scriptstyle \sin q_{y_2}(\eta \sin^2 q_x + \sin^2 q_{y_2}-\Omega^2)} &  \dfrac{\scriptstyle \cos q_{y_3}}{\scriptstyle \sin q_{y_3}(\eta \sin^2 q_x + \sin^2 q_{y_3}-\Omega^2)}\\
\scriptstyle 1 &\scriptstyle 1 &\scriptstyle 1
 \end{array}\right| =0 \ .
  \end{split}
\end{equation}

\section{Dispersion relation in the granular phononic crystal with particles possessing one translational and two rotational \textit{dofs}}\label{appendixSH}
 The determinant of the eigenvalue problem~(\ref{eqC4:detSH}) leads to
\begin{align}\label{eqC4:det1SH}
& -p \cos^2 q_x \sin^2 q_x \left( p-\Omega^2+p \cos^2 q_y +p_B p \sin^2 q_y\right) \nonumber\\
 &+ \left[p-\Omega^2+p \cos^2 q_x+ p_B p \sin^2 q_x\right] \nonumber \\
 &\times \left[-p\cos^2 q_y \sin^2 q_y + \left( -\Omega^2 + \sin^2 q_x +\sin^2 q_y\right)\left(p-\Omega^2+ p\cos^2 q_y+p_B p \sin^2 q_y\right) \right] =0 \ ,
\end{align}
which can be written in a characteristic equation for $Y=\sin^2 q_y$ 
{\small\begin{align} \label{eqC4:SySH}
& Y^2 \left[ p_B p \left(p-\Omega^2 + p \cos^2 q_x +p_B p \sin^2 q_x\right) \right]  \nonumber \\
& + Y \left[-(p-\Omega^2)(-p+\Omega^2+p_B p\Omega^2)-p(\Omega^2+p(-1+p_B \Omega^2))\cos^2 q_x - p_B p(-2p+2\Omega^2+p_Bp\Omega^2)\sin^2q_x +p^2 p_B^2 \sin^4 q_x\right]\nonumber \\
&-(p-\Omega^2+p\cos^2 q_y)(p\,\Omega^2 \cos^2q_x+(\Omega^2-\sin^2q_x)(p-\Omega^2+p_B p \sin^2 q_x))=0 \ ,
\end{align}}
and in a cubic equation for $\Omega^2$
{\small\begin{align} \label{eqC4:OmegaSH}
& -(\Omega ^2)^3\nonumber \\
&+(\Omega ^2)^2 \left[ p_B p\sin
   ^2q_x+ p_B p\sin ^2q_y+p \cos ^2q_x+p \cos
   ^2q_y+2 p+\sin ^2q_x+\sin ^2q_y\right]\nonumber \\
&-\Omega ^2 p(p_B^2 p\sin ^2q_x \sin ^2q_y+\cos ^2q_y
   \left((p_Bp+1) \sin ^2q_x+p\right)+\cos ^2q_x \left((p
   p_B+1) \sin ^2q_y+p \cos ^2q_y+p\right)\nonumber \\
 &+p_B p\sin^2q_x+p p_B \sin ^2q_y+p+2 p_B \sin ^2q_x \sin
   ^2q_y+p_B \sin ^4q_x+p_B \sin ^4q_y+2 \sin
   ^2q_x+2 \sin ^2q_y) \nonumber \\
   &+ p^2 \left(\sin ^2q_x \left(\left(p_B \sin ^2q_y+1\right)^2+\cos
   ^2q_y\right)+\left(\cos ^2q_x+1\right) \sin ^2q_y
   \left(p_B \sin ^2q_y+1\right)+p_B \sin ^4q_x
   \left(p_B \sin ^2q_y+\cos ^2q_y+1\right)\right)\nonumber \\
   &=0 \ .
\end{align}}
\section{Inexistence of SH surface waves in the absence of bending rigidity} \label{appendix:SHpB0}

When the bending rigidity parameter $p_B$ is zero, Eq.~(\ref{eqC4:SySH}) reduces to 
\begin{align}
   &Y \left[\left(p-\Omega
   ^2\right) \left(-p \sin ^2q_x+p \Omega ^2+p-\Omega ^2\right)+p \left(p
   \Omega ^2+p-\Omega ^2\right) \cos ^2q_x\right]   \nonumber \\
   &+\left(\Omega ^2-2 p\right) \left(\left(p-\Omega ^2\right) \left(\Omega ^2-\sin
   ^2q_x\right)+p \Omega ^2 \cos ^2q_x\right)=0 \ ,
   \end{align}
which leads to the wave number $q_{y_{1}}$
\begin{equation}\label{eq:qy1SH}
q_{y_1}= \arcsin \sqrt{\frac{2 \left(2 p-\Omega ^2\right) \left(\left(p-\Omega ^2\right)
   \left(\Omega ^2-\sin ^2q_x\right)+p \Omega ^2 \cos ^2q_x\right)}{p^2
   \left(3 \Omega ^2+2\right)+p \left(p \left(\Omega ^2+2\right)-2 \Omega ^2\right)
   \cos (2q_x)-2 p \left(\Omega ^2+2\right) \Omega ^2+2 \Omega
   ^4}} \ .
\end{equation}
To each frequency corresponds one wave number $q_{y_{1}}$. In absence of bending rigidity, only the boundary condition Eq.~(\ref{eqC4:BC_GX_SH1}) is applied, which leads to the equation
\begin{equation}
\begin{split} 
&(1-e^{2\text{j} q_{y_1}})-\alpha_1 (1+e^{2\text{j} q_{y_1}})=0 \ ,\\
& \Leftrightarrow \alpha_1=\frac{2 \text{j} \sin q_{y_1}}{2 \cos q_{y_1}} \ ,\\
& \Leftrightarrow \dfrac{-\text{j}p\sin q_{y_1} \cos q_{y_1}}{p( \cos^2 q_{y_1} + 1)-\Omega^2}=\frac{ \text{j} \sin q_{y_1}}{ \cos q_{y_1}} \ ,\\
& \Leftrightarrow \Omega^2=p \ .
\end{split}
\end{equation}  
This result is in accordance with the surface mode frequency Eq.~(\ref{eq:ModeS_SH}) with $p_B=0$. Nevertheless, in this case, no surface mode exists because the frequency $\Omega=\sqrt{p}$ lies in a propagation band. In fact, according to Eq.~(\ref{eq:qy1SH}), the corresponding wave number is purely real, and equal to $q_{y_1}=\pi/2$.

%

\end{document}